\documentclass[preprint,12pt,authoryear]{elsarticle}
 \usepackage{amssymb}
\pdfoutput=1
\usepackage{amsfonts,amsmath,epsfig,psfrag}
 \usepackage{amsmath}
 \usepackage{tikz}
\usepackage{pgflibraryshapes}
\usepackage{xcolor}
 \usepackage{graphicx,psfrag}
 \makeatletter \oddsidemargin 0.0cm \evensidemargin 0.0cm
\newcommand{\be}{\begin{equation}}
\newcommand{\en}{\end{equation}}
\newcommand{\la}{\label}
\newcommand{\ep}{{\varepsilon}}

\def\rr#1{(\ref{#1})}
\def\bm#1{\mbox{\boldmath{$#1$}}}
\def\ii{{\rm i}}

\begin{document}

\begin{frontmatter}
\title{{\bf Buckling and post-buckling of a compressed elastic half-space coated by double layers}}
\author[mymainaddress]{Mingtao Zhou}
\author[mymainaddress]{Zongxi Cai}
\address[mymainaddress]{Department of Mechanics, Tianjin University, Tianjin 300072, China}
\author[mysecondaryaddress]{Yibin Fu\corref{mycorrespondingauthor}}
\cortext[mycorrespondingauthor]{Corresponding author \\ {\bf \indent Dedicated to the memory of Prof. Hui-Hui Dai.} }
\address[mysecondaryaddress]{School of Computing and Mathematics, Keele University, Staffs ST5 5BG, UK}

\begin{abstract}
We investigate the buckling and post-buckling properties of a hyperelastic half-space coated by two hyperelastic layers when the composite structure is subjected to a uniaxial compression. In the case of a half-space coated with a {\it single} layer, it is known that when the shear modulus $\mu_f$ of the layer is larger than the shear modulus $\mu_s$ of the half-space, a linear analysis predicts the existence of a critical stretch and wave number, whereas a weakly nonlinear analysis predicts the existence of a threshold value of the modulus ratio $\mu_s/\mu_f\approx 0.57$ below which the buckling is super-critical and above which the buckling is sub-critical. It is shown in this paper that when another layer is added, a larger variety of behaviour can be observed. For instance, buckling can occur at a preferred wavenumber super-critically even if both layers are softer than the half-space although the top layer would need to be harder than the bottom layer. When the shear modulus of the bottom layer lies in a certain interval, the super-critical to sub-critical transition can happen a number of times as the shear modulus of the top layer is increased gradually. Thus, an extra layer imparts more flexibility in producing wrinkling patterns with desired properties, and our weakly nonlinear analysis provides a road map on the parameter regimes where this can be achieved.
\end{abstract}

\end{frontmatter}

\section{Introduction}
\setcounter{equation}{0}
Buckling of a coated elastic half-space induced by uni-axial compression is a topic that has been much studied in recent decades.
Motivated mostly by the desire to suppress buckling as a precursor to structural failure, early studies under the framework of nonlinear elasticity include the linear analyses by \citet{dn1980}, \citet{sks1994},  \citet{os1996}, \citet{bon1997},   \citet{so1997},  and the nonlinear post-buckling analysis by   \citet{cai-fu1999}. The subsequent experimental work by  \citet{boden1998, boden1999} on pattern formation at the micrometer and sub-micrometer scales opened the possibility to use such pattern formation to achieve a variety of useful purposes. Known applications now range from cell patterning \citep{chien2012}, optical gratings \citep{lee2010, ma2013, kim2013}, creation of surfaces with desired wetting and adhesion \citep{chan2008, zhang-gao2012}, to buckling-based metrology \citep{stafford2004}. Pattern formation is also known to play an important role in many biological processes where the main driving mechanism is growth \citep{goriely2017}. Motivated by these newly found applications, further linear and nonlinear analyses have been conducted by   \citet{cai-fu2000},   \citet{so2002},   \citet{sx2012},   \citet{hutchinson2013},   \citet{Ci2014},   \citet{FC2014}, \citet{CF2015},  \citet{zhang2017},  \citet{holland2017},   \citet{Alawiye2019},   \citet{cai-fu2019},   \citet{Alawiye2020}, and   \citet{rd2021}. There also exists a large body of literature that employs  approximate plate theories  and simplified interfacial conditions in order to provide qualitative understanding of the phenomena observed; see, e.g.,   \citet{CH2004},  \citet{HH2005},   \citet{AB2008}, and   \citet{SJ2008}.  We refer to the reviews by   \citet{gg2006},   \citet{yang2010},  \citet{li-cao2012},  and   \citet{dw2020}  for a comprehensive review of the relevant literature.

A major result arising from a weakly nonlinear analysis  \citep{cai-fu1999, hutchinson2013, Alawiye2020} is that there exists a critical stiffness ratio that marks the transition from subcritical bifurcation to supercritical bifurcation. It can then be expected, and indeed confirmed by many recent numerical and experimental studies, that in the subcritical regime localization is the norm \citep{cai2012} whereas in the supercritical regime stable periodic patterns and further secondary bifurcations are the norm \citep{brau2011, ch2012a, liu2015, fu-cai2015, budday2015, zhao-cao2015, zhuo-zhang2015b, zhuo-zhang2015a, cai-fu2019, cx2020}.

In this paper, we extend the analysis of  \citet{cai-fu1999} and  consider the effects of adding an extra layer to the film-substrate bilayer structure.  The structure of a substrate coated by two layers has been recognized to offer enhanced capabilities in buckling based metrology \citep{ncr2006, jcl2012, ljc2016},
stretchable electronics \citep{czh2014}, and surface pattern switching \citep{jcl2012, wzn2020}. Multi-laying is of course the norm in biological systems (e.g. the human skin), whereas in other situations adding extra layers arises out of necessity. For instance, in cell patterning, there may exist the conflicting demands that on the one hand well-ordered wrinkling patterns are desired, and on the other hand the top film is required to be sufficiently soft to dictate a certain cell behavior.

For the structure under consideration, there exist two modulus ratios, namely $r_1=\mu_s/\mu_1$ and $r_2=u_s/\mu_2$, where $\mu_s$, $\mu_1$ and $\mu_2$ are the shear moduli of the substrate, the first layer (i.e. the lower layer), and the top layer, respectively. In this case, the sign of the nonlinear coefficient $c_1$ in the amplitude equation, that determines the subcritical to supercritical transition, is a function of $r_1$ and $r_2$. Our main objective is to determine the dividing curve in the $(r_1, r_2)$-plane where this coefficient vanishes. It is hoped that such information can help guide the design process when robust wrinkles are desired. We present results for three representative cases when the thickness ratio of the two layers is equal to $1$, $0.1$ and $10$, respectively, but the methodology is valid for any other thickness ratio (and for any material model).

The rest of this paper is divided into five sections as follows. After formulating the buckling problem in the next section, we present in Section 3 the necessary linear analysis that produces the bifurcation condition. The bifurcation condition exhibits two interesting features. Firstly, even if both layers are softer than the half-space, the stretch can still exhibit a maximum at a finite wavenumber. Secondly, there is a range of thickness ratios for which multiple stretch maxima exist and mode switching becomes possible as a material parameter is varied. This is in contrast with the situation of a single layer where a stretch maximum is only possible when the layer is stiffer than the half-space and no mode switching is possible. In Section 4 we conduct a weakly nonlinear analysis in order to determine the nature of buckling, namely whether it is super-critical or sub-critical. A richer variety of behaviour than what is possible for a single layer is uncovered. The paper is concluded in Section 5 with a summary, and a discussion of limitations of the current study, and possible future work.

\section{Governing equations}
We first formulate the governing equations for a general homogeneous elastic body $B$ composed of a non-heat-conducting incompressible elastic material. Three configurations of $B$ are involved in our analysis: the initial unstressed configuration $B_0$, the finitely stressed equilibrium configuration denoted by $B_{\rm e}$, and the current configuration $B_t$ that is obtained from $B_{\rm e}$ by a small perturbation. The position vectors of a representative material particle relative to a common Cartesian coordinate system are denoted by ${\bm X}$, ${\bm x}$, and  $\tilde{\bm x}$ in $B_0$, $B_{\rm e}$ and $B_t$, respectively. The associated coordinates are written as
$X_A$, $x_i$ and $\tilde{x}_i$. We write
\be \tilde{\bm x} = {\bm x}({\bm X}) + {\bm u}({\bm x}), \la{coe21} \en
where ${\bm u}({\bm x})$ is the incremental displacement associated with the deformation $B_{\rm e} \to B_t$.

The deformation gradients arising from the deformations $B_0 \to B_t$ and $B_0 \to B_{\rm e}$ are denoted by $\tilde{\bm F}$ and $\bm{\bar{F}}$, respectively, and are defined by
their Cartesian components \be \tilde{F}_{iA} = \frac{\partial{\tilde{x}_i}} {\partial{X_A}}, \;\;\;\; \bar{F}_{iA} = \frac{\partial{x_i}} {\partial{X_A}}. \la{dg21} \en
It then follows that
\be \tilde{F}_{iA} = (\delta_{ij}+u_{i,j}) \bar{F}_{jA}, \la{dg22} \en
where here and henceforth a comma signifies differentiation with respect to the indicated coordinate, with \lq$, A$' and \lq $, j$' meaning differentiation with respect to $X_A$ and $x_j$, respectively.

%

In the absence of body forces, the equations of equilibrium and the incompressibility constraint are given by
\be \pi_{iA,A} = 0, \;\;\;\; \textup{det} \bm{F} = 1,\la{mon21} \en
where $\bm{\pi}$ is the first Piola-Kirchhoff stress which, in component form, is given by
\be \pi_{iA} = \frac{\partial W} {\partial F_{iA}} - pF^{-1}_{Ai}, \la{pks21} \en
with $W$ denoting the strain-energy function (per unit volume in the reference configuration) and $p$ a Lagrange multiplier enforcing the incompressibility constraint. We denote by $\bar{p}$ and $\bar{p}+\tilde{p}$ the values of $p$ associated with the deformations $B_0 \to B_{\rm e}$ and $B_0 \to B_t$, respectively. Thus $\tilde{p}$ is the increment of $p$ which is an additional field induced by the constraint of incompressibility.

We define the incremental stress tensor ${\bm \chi}$ through
\be {\bm \chi}=\bar{J}^{-1} (\tilde{\bm \pi}-\bar{\bm \pi}) \bar{F}^T, \la{chi} \en
where $\bar{J}={\rm det}\, \bar{F}$, and $\bar{\bm \pi}$ and $\tilde{\bm \pi}$ are the first Piola-Kirchhoff stresses in $B_{\rm e}$ and $B_t$, respectively.
It can be shown, see e.g. \citet{fr1994}, that ${\bm \chi}$ satisfies the equilibrium equation
\be \chi_{ij,j} = 0, \la{mon22} \en
and has a Taylor expansion given by
$$ \chi_{ij} = \mathcal{A}^{1}_{jilk} u_{k,l} + \frac{1}{2} \mathcal{A}^{2}_{jilknm} u_{k,l} u_{m,n} + \frac{1}{6} \mathcal{A}^{3}_{jilknmqp} u_{k,l} u_{m,n} u_{p,q}$$ \be
+\bar{p} (u_{j,i} - u_{j,k} u_{k,i} + u_{j,k} u_{k,l} u_{l,i}) - \tilde{p} (\delta_{ji} - u_{j,i} + u_{j,k} u_{k,i}) + O(\epsilon^4), \la{tf22} \en
where $\epsilon$ is a small parameter characterizing the magnitude of $u_{i,j}$ and $\tilde{p}$, and the tensors $\bm{\mathcal{A}}^{1}$, $\bm{\mathcal{A}}^{2}$ and $\bm{\mathcal{A}}^{3}$ are the first-, second- and third-order tensors of instantaneous elastic moduli in $B_{\rm e}$ whose expressions are typified by \citep{co1971}
\be \mathcal{A}^{1}_{jilk} = \bar{J}^{-1} \bar{F}_{jA} \bar{F}_{lB} \frac{\partial^2 W}{\partial F_{iA} \partial F_{kB}} \bigg|_{\bm{F} = \bm{\bar{F}}}. \la{iem21} \en
On substituting \rr{tf22} into \rr{mon22} and simplifying with the use of the identity $(J F^{-1}_{Ai})_{,A} \equiv 0$, we obtain
$$ \mathcal{A}^{1}_{jilk} u_{k,lj} + \mathcal{A}^{2}_{jilknm} u_{m,n} u_{k,lj} + \frac{1}{2} \mathcal{A}^{3}_{jilknmqp} u_{m,n} u_{p,q} u_{k,lj}$$ \be
 - \tilde{p}_{,j} (\delta_{ji} - u_{j,i} + u_{j,m} u_{m,i}) + O(\epsilon^4) = 0. \la{mon23} \en
These incremental equilibrium equations are supplemented by the incompressibility condition in the form
\be u_{i,i} = \frac{1}{2} u_{m,n} u_{n,m} - \frac{1}{2} (u_{i,i})^2 - \textup{det} (u_{m,n}). \la{dpt21} \en
We now specialize the above equations to the structure of an incompressible elastic half-space coated by two incompressible elastic layers. Each component (layer or half-space) in this structure is a homogeneous elastic body to which the above equations apply. We choose our common coordinate system such that the half-space, the first layer, and second layer are defined by $-\infty < x_2 \leq 0$, $0\leq x_2 \leq h_1$, $h_1\leq x_2 \leq h_1+h_2$, respectively, where $h_1$ and $h_2$ are constant thicknesses of the two layers to be specified.

To simplify analysis, we assume that the bonded structure is in a state of plane strain so that $u_3 = 0$ and $u_1$ and $u_2$ are independent of $x_3$. We also assume that the principal axes of stretch corresponding to the finite deformation are aligned with the coordinate axes.

Our problem is then to solve \rr{dpt21} and \rr{mon23} in $-\infty < x_2 \leq h_1+h_2$ subject to the following auxiliary conditions:\medskip

\noindent $(\textup{i})$ Traction-free boundary conditions,
\be {\chi}_{i2} = 0, \phantom{aaa} \textup{on} \phantom{a} x_2 = h_1 + h_2, \la{bc21}\en
$(\textup{ii})$ interfacial continuity conditions,
\be  [u_i] = 0, \phantom{aaa} [\chi_{i2}]=0, \phantom{aaa} \textup{at} \phantom{a} x_2 = 0, \;{\rm or}\; h_1, \la{icadded} \en
$(\textup{iii})$ and decay conditions,
\be u_i \to 0 \phantom{aa} \textup{as} \phantom{a} x_2 \to -\infty, \la{dc21} \en
where the notation $[f]$ denotes the jump of $f$ when an interface is crossed.
The trivial solution $u_i=0$, $\tilde{p}=0$ is clearly one solution. Our aim is to find the conditions under which the above boundary value problem has a non-trivial solution.

Our numerical calculations will be carried out for the case when the layers and half-space are all composed of either neo-Hookean or Gent materials, and the prestress in $B_{\rm e}$ takes the form of a uniaxial compression along the $x_1$-direction with stretch $\lambda$. For neo-Hookean and Gent materials, the strain energy function is given by
\be W = \frac{1}{2} \mu (\textup{tr} \bm{B} - 3), \;\;\;\;{\rm and}\;\; W=-\frac{1}{2} \mu J_m \ln \left(1-\frac{I_1-3}{J_m} \right), \la{ef21} \en
respectively, where $I_1={\rm tr}\, {\bm B}$, $\bm{B}=\bm{F}\bm{F}^{\textup{T}}$,   $\mu$ is the shear modulus, and $J_m$ is a constitutive constant characterizing material extensibility with the limit $J_m \to \infty$ recovering the neo-Hookean model. We denote the shear moduli of the substrate and the two layers by $\mu_s$, $\mu_1$ and $\mu_2$, respectively, and define the dimensionless parameters
 $r_1=\mu_s/\mu_1$, $r_2=\mu_s/\mu_2$. When the Gent material model is used, we shall assume that $J_m$ takes the same value in all the three components, but our methodology can accommodate different values of $J_m$ and is in fact valid for any strain-energy function. All of our symbolic manipulations and numerical integrations are carried out with the aid of Mathematica \citep{wolf2019}.

%

\section{Linear theory}

In preparation for the later nonlinear analysis, we first consider the linearized version of the problem specified by \rr{mon23}, \rr{dpt21}, and \rr{bc21}$-$\rr{dc21}. The linearized governing equations are
\begin{equation}
u_{i,i} = 0, \phantom{aa}
\mathcal{A}^{1}_{jilk} u_{k,lj} - \tilde{p}_{,i} = 0,\phantom{aa}
-\infty < x_2 \leq h_1+h_2,\la{lge33}
\end{equation}
and the auxiliary conditions are
\begin{equation}
{T}^{(l)}_{i} = 0 \phantom{aa}
\textup{on} \phantom{a} x_2 = h_1+h_2,\la{ac31}
\end{equation}
\begin{equation}
[{u}_i] = 0, \phantom{aaa}
[{T}^{(l)}_{i}] = 0, \phantom{aaa}
\textup{on} \phantom{a} x_2 =0, \; {\rm or}\; h_1,\la{ac32}
\end{equation}
\begin{equation}
u_i \to 0 \phantom{aa} \textup{as} \phantom{a} x_2 \to -\infty,\la{ac34}
\end{equation}
where the linearized traction $T^{(l)}_i$ is given by
\begin{equation}
T^{(l)}_i = \mathcal{A}^{1}_{2ilk} u_{k,l} + \bar{p} u_{2,i} - \tilde{p} \delta_{2i}.
\nonumber
\end{equation}
Equation \rr{lge33}$_{1}$ implies the existence of a \lq stream' function $\psi (x_1, x_2)$ such that
\begin{equation}
u_1 = \psi_{,2}, \phantom{aaa}
u_2 =  - \psi_{,1}.\la{sf31}
\end{equation}
Substituting \rr{sf31} into \rr{lge33}$_{2}$ and eliminating $\tilde{p}$ through cross differentiation, we obtain \citep{do1991}
\begin{equation}
\gamma \psi_{,2222} + 2 \beta \psi_{,1122} + \alpha \psi_{,1111} = 0,\la{ge31}
\end{equation}
where
\be
\alpha =\mathcal{A}^{1}_{1212}, \;\;\;\; 2 \beta =\mathcal{A}^{1}_{1111}+\mathcal{A}^{1}_{2222}-2 \mathcal{A}^{1}_{1122}-
2 \mathcal{A}^{1}_{1221}, \;\;\;\; \gamma=\mathcal{A}^{1}_{2121}. \la{abc} \en
We look for a periodic wrinkling solution in the form
\begin{equation}
\psi = H(kx_2) \textup{e}^{\textup{i} k x_1},\la{tws31}
\end{equation}
where $k$ is the wave number. On inserting \rr{tws31} into \rr{ge31} and solving the resulting fourth-order ordinary differential equation for $H$, we obtain
\begin{equation}
H(kx_2) = \left\{ \begin{array}{ll} A_1 \phantom{.} \textup{exp} (k s_1 x_2) + A_2 \phantom{.} \textup{exp} (k s_2 x_2), & x_2 \in (-\infty, 0), \\
\sum_{j=1}^{4} \hat{A}_{j} \phantom{.} \textup{exp} (k s_j x_2), & x_2 \in (0, h_1), \\
\sum_{j=1}^{4} \tilde{A}_{j} \phantom{.} \textup{exp} (k s_j x_2), & x_2 \in (h_1, h_1+h_2), \end{array} \right.
\la{an31}
\end{equation}
where $A_1, A_2, ..., \tilde{A}_3, \tilde{A}_4$ are disposable constants, and $s_1$ and $s_2$ are the two roots of
\begin{equation}
\gamma s^4 - 2 \beta s^2 + \alpha = 0,\la{eqn31}
\end{equation}
that have positive real parts (so that $H\to 0$ as $x_2\to -\infty$), and $s_3=-s_1$, $s_4=-s_2$. In the above expressions, we have assumed that $s_1$ and $s_2$ are independent of the elastic moduli so that they take the same values in the layers and half-space. For the neo-Hookean material model, we have $s_1=1, s_2=\lambda^2$.

On substituting the above solutions into the auxiliary conditions \rr{ac31}$-$\rr{ac34}, we obtain a matrix equation of the form $K {\bm c}={\bm 0}$, where $K$ is a $10\times 10$ matrix and ${\bm c}$ is the column vector formed from the 10 unknowns $A_1, A_2,  ...,\tilde{A}_3, \tilde{A}_{4}$. The existence of a non-trivial solution requires
\be
{\rm det}\,K=0, \la{bif} \en which is the bifurcation condition relating the pre-stretch $\lambda$ to the wavenumber $k$. The wavenumber appears in the bifurcation condition through $kh_1$ and $kh_2$. Thus, in terms of the thickness ratio $h=h_2/h_1$, the bifurcation condition \rr{bif} takes the form $F(kh_1, h, \lambda, r_1, r_2)=0$.

For each choice of $h$, $r_1$ and $r_2$, the above bifurcation condition define $\lambda$ as a function of $kh_1$. It has two important limits, $k\to 0$ or $k \to \infty$, for which the corresponding value of $\lambda$ is well-known when the material is neo-Hookean. When $k \to 0$, which is equivalent to $h_1 \to 0$ and $h_2 \to 0$, the effect of the two layers becomes negligible and the critical value of $\lambda$ is that associated with a half-space, that is $0.54$ as first obtained by   \citet{bi1963}. When $k \to \infty$, the wavelength of the modes tends to zero and the critical modes will localize either near the traction-free surface or at one of the two interfaces (between the two layers or between the first layer and the half-space). The mode localized near the traction-free surface is again a surface wave mode with critical value of $\lambda$  equal to $0.54$, and this mode corresponds to the same branch/curve that tends to $0.54$ when $k \to 0$.
The interfacial modes have previously been studied by  \citet{do1991} and    \citet{fu2005}. It was shown in \citet{fu2005} that the {\it surface impedance tensor} $M$ for a pre-stressed half-space occupying the region $x_2>0$, with principal axes of stretched aligned with the coordinate axes, has the explicit expression
\be M=\left(\begin{array}{cc} M_1 & \ii M_4 \\ -\ii M_4 & M_2 \end{array} \right), \la{mm} \en
with
$$ M_1=\sqrt{T_{11} (Q_{11}+T_{22}+2 \sqrt{Q_{22} T_{11}}-2R_{12}-2 R_{21})}, $$ $$
M_2=\frac{M_1^3}{2 T_{11}^2}-\frac{M_1}{2 T_{11}} (Q_{11}+T_{22}-2 R_{12}-2 R_{21}), $$
$$ M_4=\frac{M_1^2}{2 T_{11}}-\frac{1}{2} (Q_{11}+T_{22}-2 R_{12}+2 \bar{p}), $$
and
$$ T_{ij}={\cal A}^{1}_{2i2j}, \;\;\;\; R_{ij}={\cal A}^{1}_{1i2j}, \;\;\;\;Q_{ij}={\cal A}^{1}_{1i1j}. $$
\begin{figure}[h]
\begin{center}
\begin{tabular}{cc}
\includegraphics[scale=0.35]{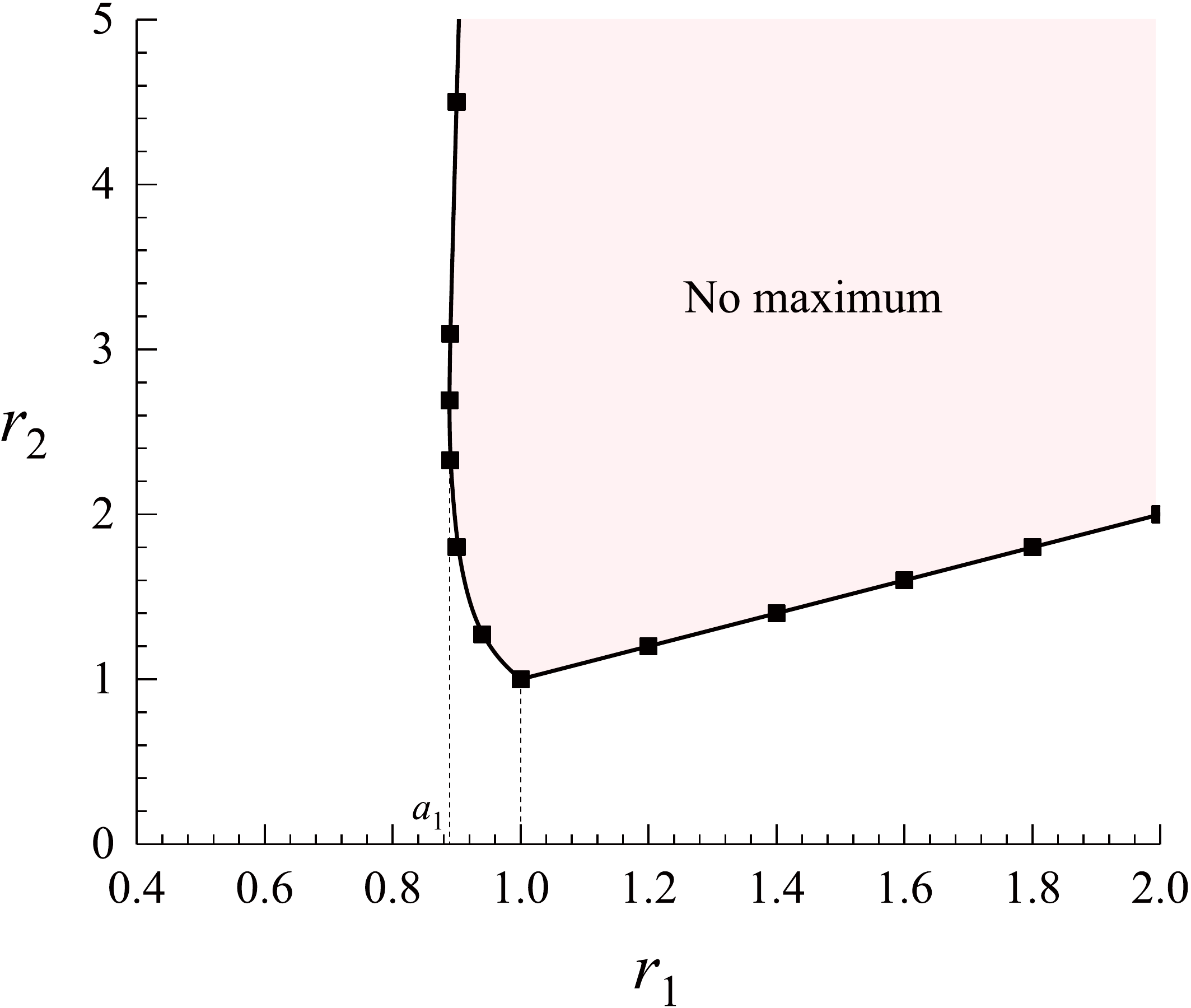}&\includegraphics[scale=0.35]{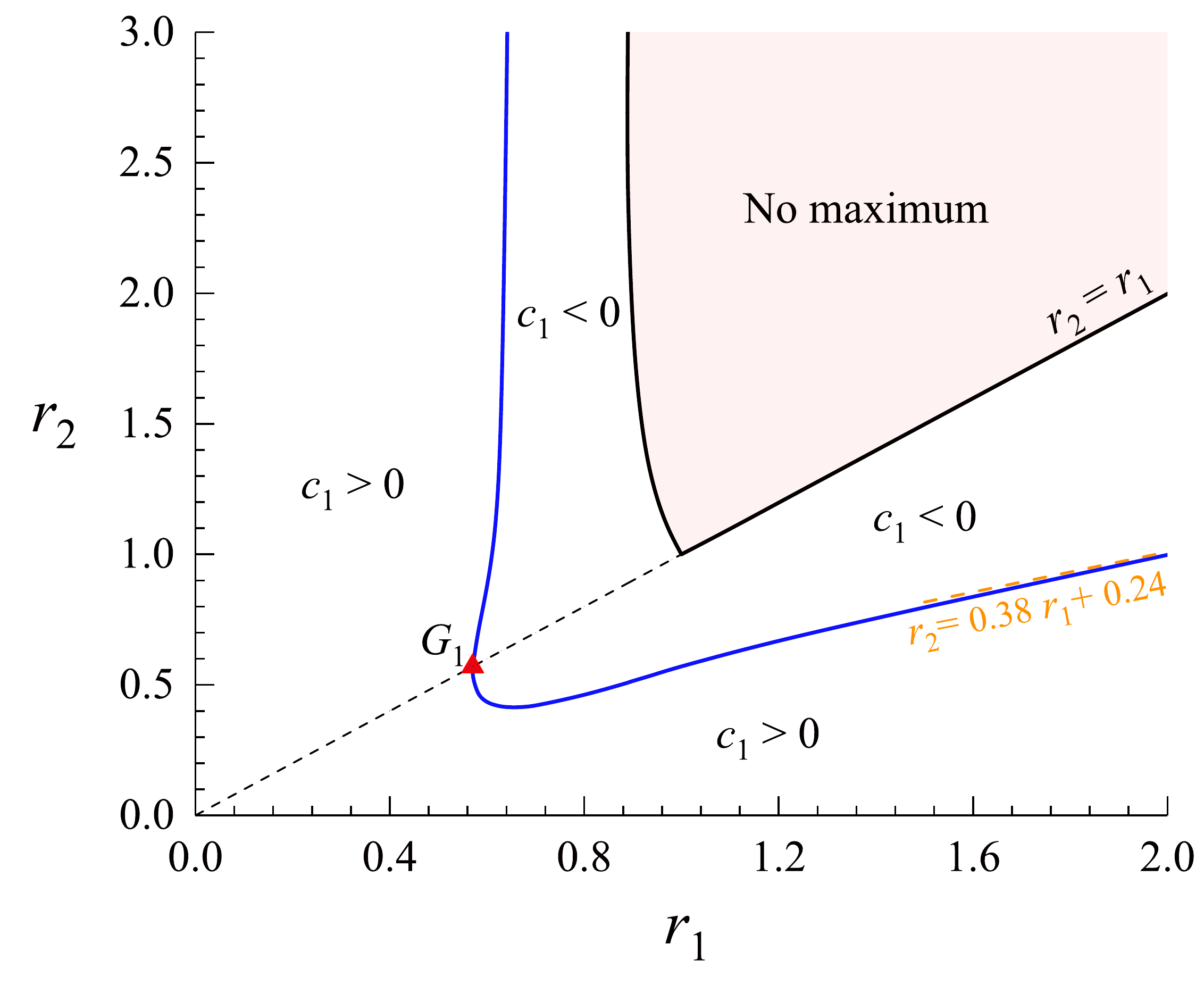}
\end{tabular}
\caption{Linear and nonlinear bifurcation behavior for $h=1$. (a) Parameter domain (unshaded) where the bifurcation stretch has a maximum. The left-most point of the bounding curve has coordinates $(a_1, 2.69)$ where $a_1 = 0.89$. (b) Dependence of the sign of $c_1$ on $r_1$ and $r_2$ with $G_1$ having coordinates $(0.57, 0.57)$ and the blue line approaching $r_1=0.57$ as $r_2 \to \infty$.}
\label{h1}
\end{center}
\end{figure}
\begin{figure}[h]
\begin{center}
 \includegraphics[width=.99\textwidth]{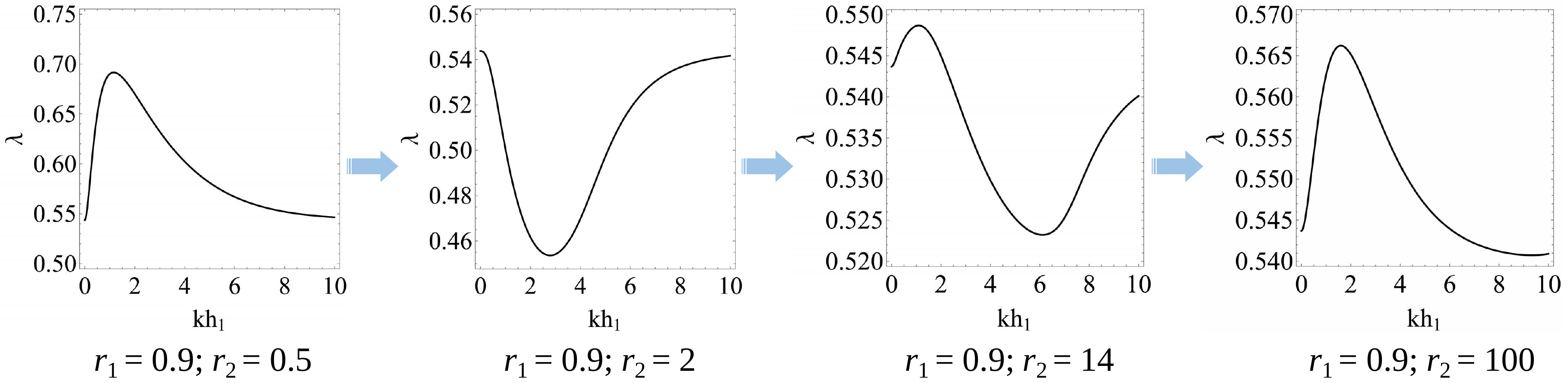}
\caption{Evolution of the bifurcation curve when $h=1, \,r_1=0.9$ showing the fact that as $r_2$ is increased, the stretch maximum first disappears and then re-merges at a larger value of $r_2$.}
\label{evo1}
\end{center}
\end{figure}
\noindent The bifurcation condition for a single half-space may be written as ${\rm det}\,M=0$ that reproduces the well-known equation $\lambda^6+\lambda^4+3 \lambda^2-1=0$ for the neo-Hookean material. This condition yields the Biot value $\lambda=0.54$. The bifurcation condition for an interfacial mode localized at the interface of two half-spaces is given by ${\rm det}\,(M+\hat{M}^*)=0$, where $M$ has the same meaning as above, $\hat{M}$ can be computed using \rr{mm} but with material constants replaced by those for the other half-space, and the superscript \lq\lq *" signifies complex conjugation. With the ratio of the two shear moduli denoted by $r$ such that $0 \le r \le 1$ (if $r$ is larger than $1$ we use its inverse instead since the two half-spaces can be exchanged), the bifurcation condition ${\rm det}\,(M+\hat{M}^*)=0$ for the case of uni-axial compression reduces to
\be (r^2+1) (\lambda^6+\lambda^4+3 \lambda^2-1)+2 r (\lambda^6+3 \lambda^4-2 \lambda^2+1)=0; \la{intmm} \en
see also \citet{do1991}. It is known that a necessary condition for ${\rm det}\,(M+\hat{M}^*)=0$ to be satisfied is that either ${\rm det}\,M<0$ or
${\rm det}\, \hat{M}^*<0$, that is the interfacial mode occurs at a value of $\lambda$ smaller than the Biot value $0.54$.  The solution of \rr{intmm} for $\lambda$ as a function of $r$ is a monotonically decreasing function of $r$, with maximum and minimum given by $0.54$ and $0$, respectively. It is found that only one interfacial mode can exist as $k \to \infty$ and this mode is always localized at the interface with greater contrast in stiffness. Since this mode lies below the curve that tends to $0.54$ when $k \to 0$ or $\infty$, it is of little interest in the current context and hence will not be displayed. However, the two limits discussed above are used as useful checks on our numerical results for intermediate values of $k$.

The bifurcation condition discussed above is next used to generate plots in the $(r_1, r_2)$-plane showing domains where a stretch maximum exists and where mode switching takes place. By mode switching we refer to situations where at a particular set of material parameters two equal maxima of $\lambda$ occur at two different values of $k$ (corresponding to two modes with short and long wavelengths, respectively), and
a small perturbation of such parameter values can make one or the other the only preferred mode. In other words, with a slight change of the material parameters, the bifurcation mode can switch from a short mode to a long mode or vice versa \citep{jcl2012}.
We shall present illustrative results for three representative cases, and for the neo-Hookean material model. Before presenting results for each case, however, we first observe some general features shared by all cases. These general features correspond to the three limits, $r_1 \to 1$, $r_2 \to r_1$, or $r_2 \to \infty$,  under which the structure under consideration reduces to a half-space coated by a single layer. Since for the latter reduced case a stretch maximum exists only when the single layer is stiffer than the substrate, the boundary of domain where a stretch maximum exists for the current problem must contain the point $(r_1, r_2)=(1, 1)$ and must approach the asymptote $r_1=1$ as $r_2 \to \infty$. It may also be deduced that the semi-infinite straight line $r_2 = r_1$ with $r_1>1$ must lie in the domain of non-existence of a stretch maximum. It turns out that this line is actually part of the boundary of this domain.

\begin{figure}[h]
\begin{center}
\begin{tabular}{cc}
\includegraphics[scale=0.35]{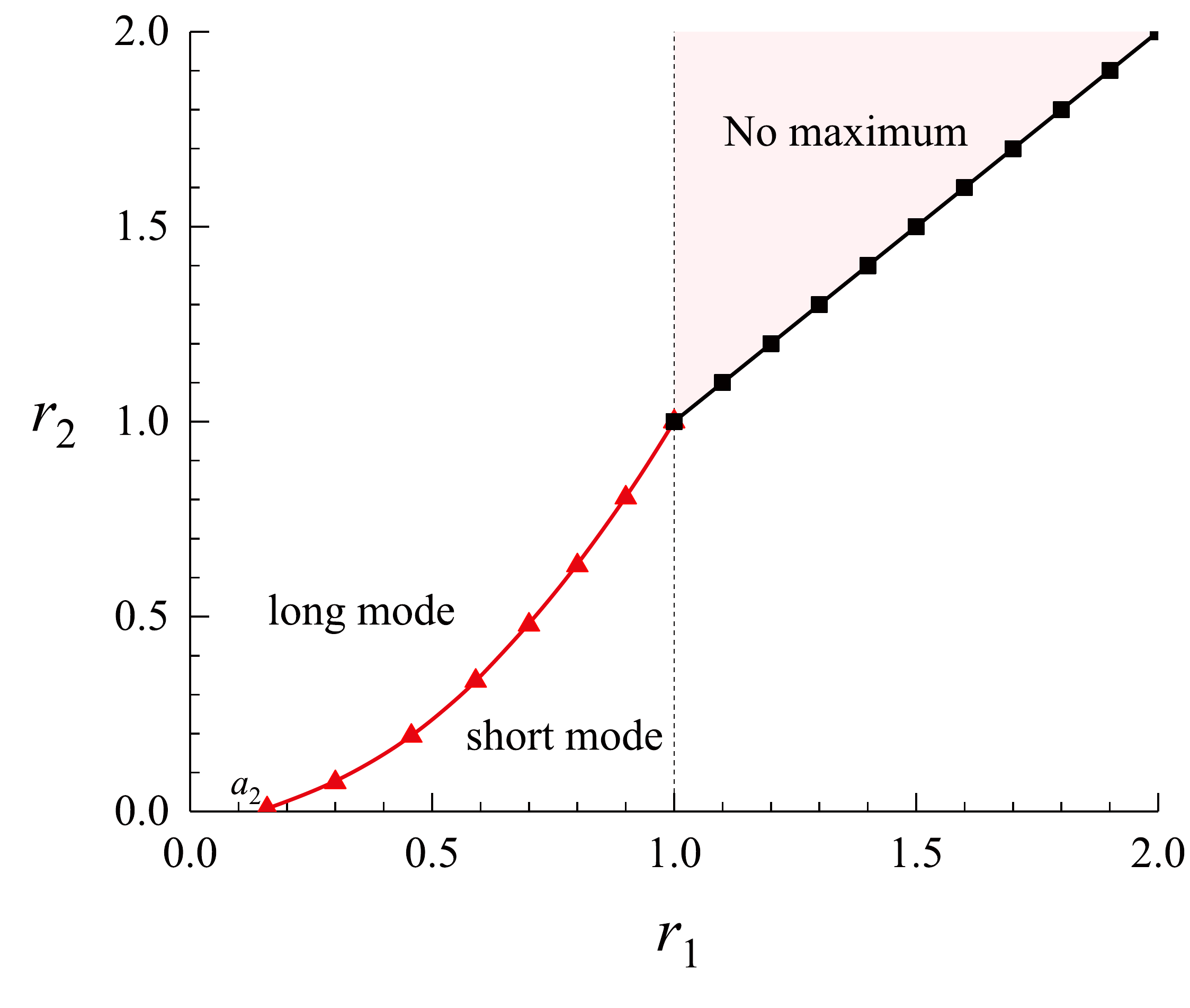}&\includegraphics[scale=0.35]{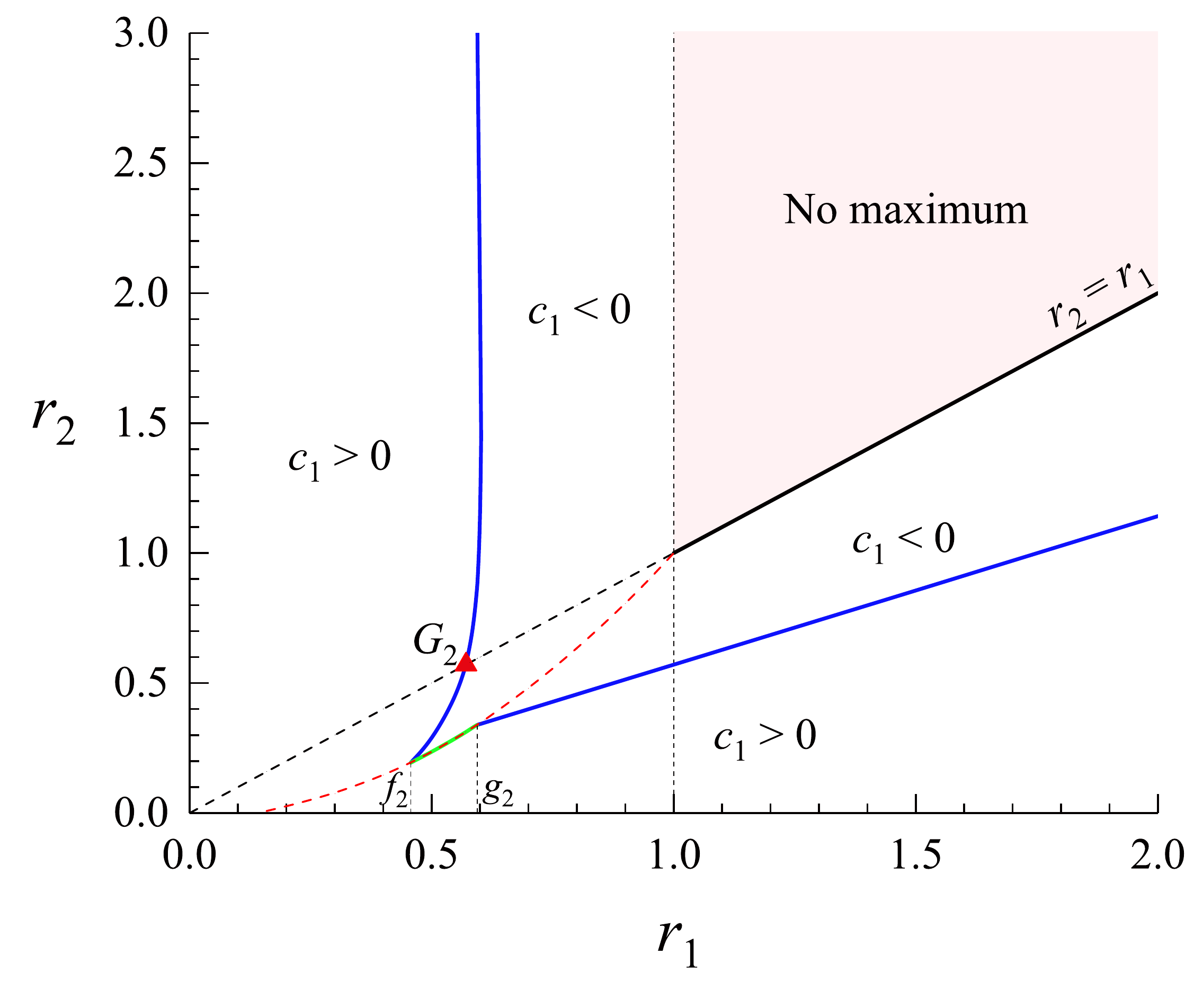}
\end{tabular}
\caption{Results for $h=0.1$. (a) Parameter domain (unshaded) where the bifurcation stretch has a maximum.
The red line is where mode switching takes place and it terminates at $(a_2, 0.008)$ where $a_2=0.16$. (b) Dependence of the sign of $c_1$ on $r_1$ and $r_2$ with $G_2$ having coordinates $(0.57, 0.57)$ and the blue line approaching $r_1=0.57$ as $r_2 \to \infty$. The dashed red curve is where mode switching takes place, and the green line coinciding with the red dashed line is where the sign change of $c_1$ is due to mode jumping, instead of via $c_1=0$ (i.e. $c_1\ne 0$ on the green line except at the two ends). The values of $r_2$ at the ends of the dotted vertical lines starting at $f_2=0.46$ and $g_2=0.59$ are $0.19$ and $0.34$, respectively.}
\label{h01}
\end{center}
\end{figure}

In Figs\,\ref{h1}(a), \ref{h01}(a) and \ref{h10}(a), we have shown the domain in the $(r_1, r_2)$-plane where the bifurcation condition gives a stretch maximum for the three representative cases $h=1, 0.1$ or $10$, respectively (unshaded region). These numerical results are all consistent with the general observation made above. Additionally, with regard to Figs\,\ref{h1}(a), we note that for each fixed $r_1$ in the interval $(a_1, 1)$, as $r_2$ is increased from zero, a stretch maximum first exists, then disappears, and finally emerges again at a large value of $r_2$. This behaviour is displayed in Fig.\,\ref{evo1} for $r_1=0.9$. Similar behaviour is observed for fixed $r_2>1$ and variable $r_1$. Another important result seen in Fig.\,\ref{h1}(a) is that a stretch maximum exists for all values of $r_2$ if $r_1<a_1$, or for all values of $r_1$ if $r_2<1$. The former scenario corresponds to the fact that provided the lower layer is sufficiently harder than the substrate (more precisely $\mu_1> a_1^{-1} \mu_s$), a stretch maximum exists no matter how soft the top layer is. Also, when both layers are softer than the half-space, a stretch maximum is still possible provided the first layer is softer than the second layer (i.e. $r_1>r_2$).

\begin{figure}[h]
\begin{center}
 \includegraphics[width=.99\textwidth]{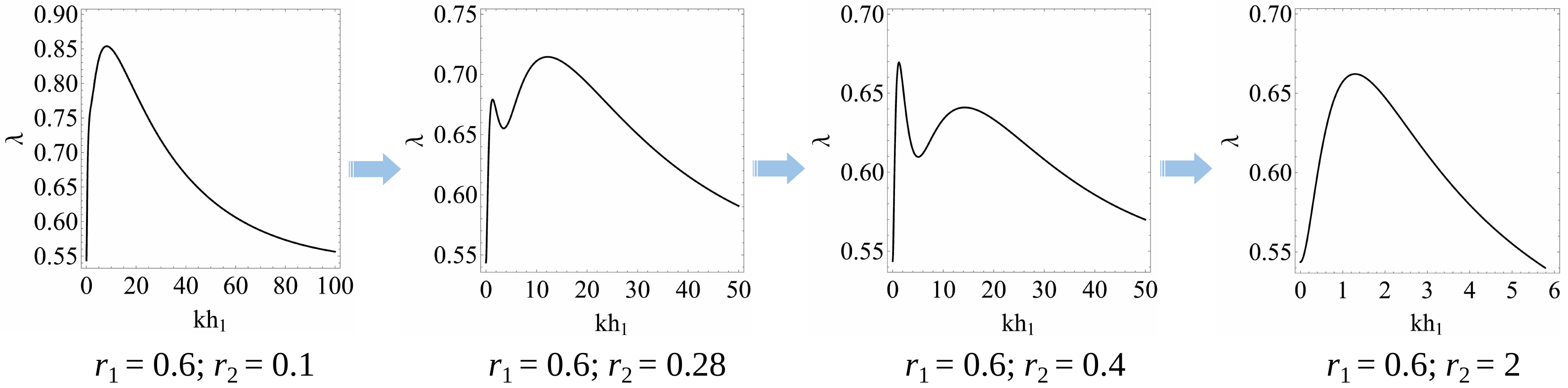}
\caption{Evolution of the bifurcation curve when $h=0.1,\, r_1=0.6$ showing the fact that as $r_2$ is increased, mode switching may take place once.}
\label{evo2}
\end{center}
\end{figure}

As $h$ is varied around $1$, it is found that multiple stretch maxima may occur when $|h-1|$ is sufficiently large.  Fig.\,\ref{h01}(a) shows the counterpart of Fig.\,\ref{h1}(a) when $h=0.1$ (the lower layer is now 10 times as thick as the upper layer). The left boundary of the shaded domain is not a straight line although it looks
that way. It actually has a similar shape to its counterpart in Fig.\,\ref{h1}(a) but the $r_1$ in the current case varies in a smaller interval, namely between $0.998$ and $1$.
The red line is where mode switching takes place (short modes are preferred below the line and long modes are preferred above the line). A typical example of mode switching is displayed in Fig.\,\ref{evo2} for $r_1=0.6$.

\begin{figure}[h]
\begin{center}
 \begin{tabular}{cc}
\includegraphics[scale=0.35]{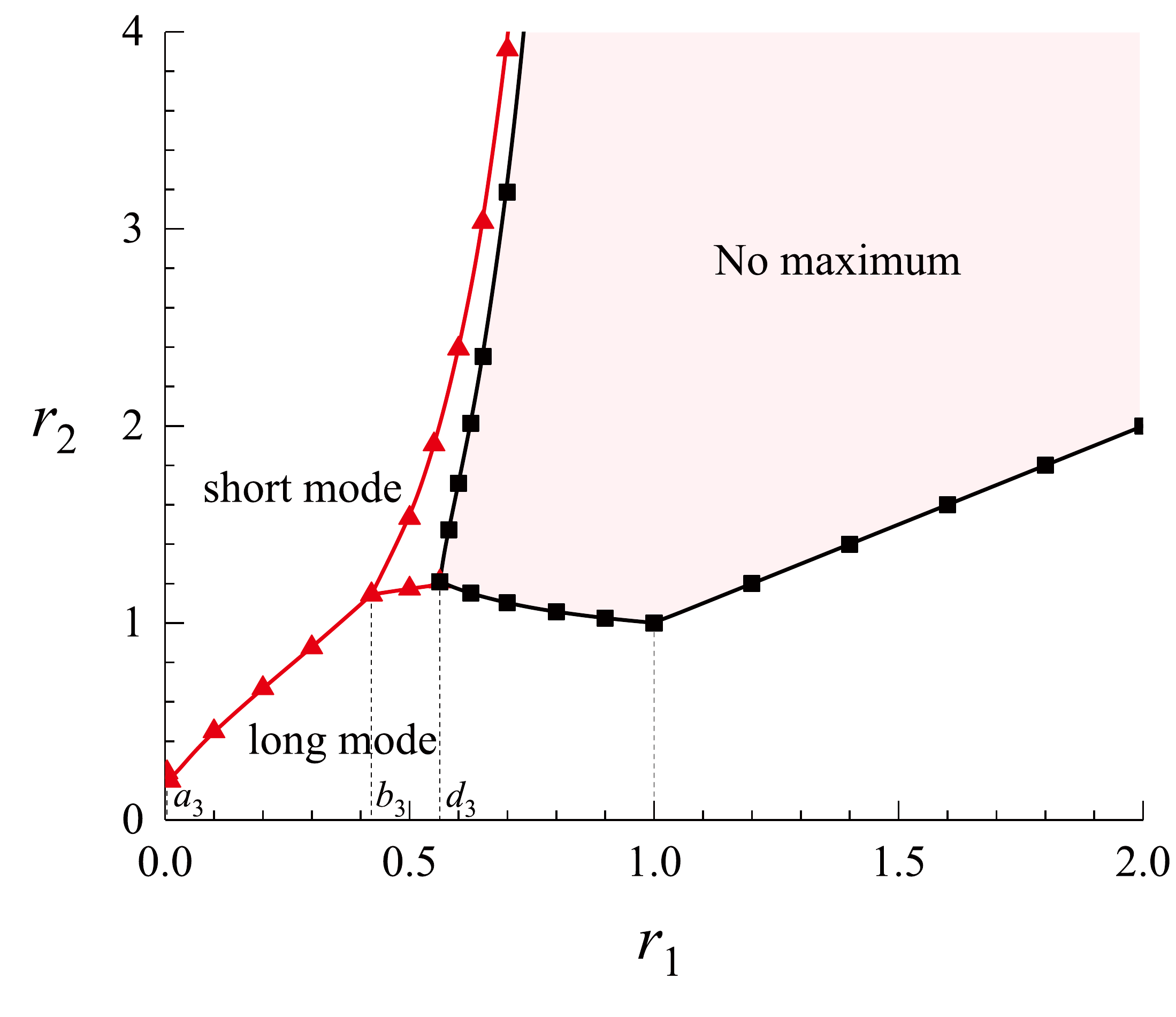}&\includegraphics[scale=0.35]{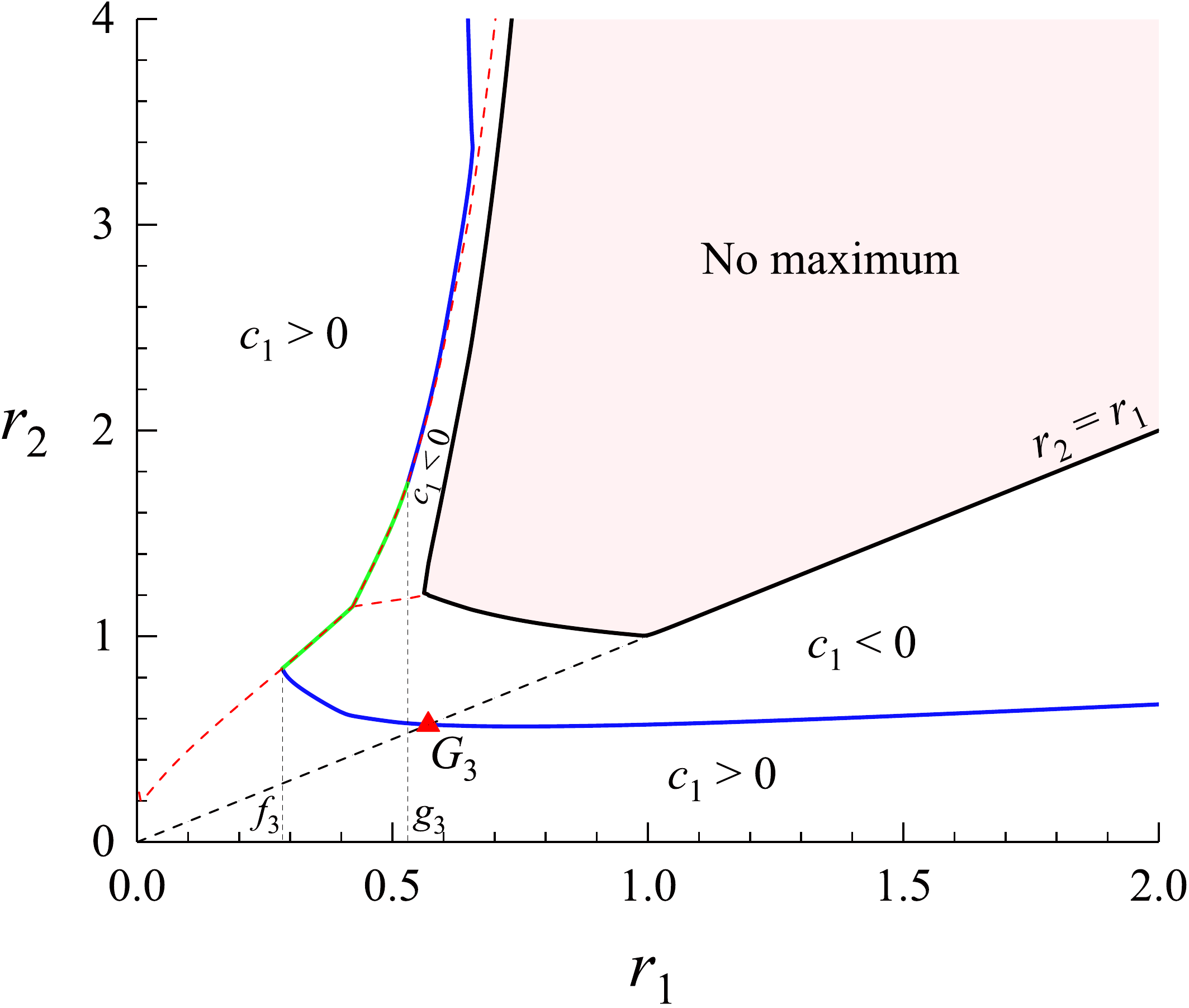}
\end{tabular}
\caption{Results for $h=10$. (a) Parameter domain (unshaded) where the bifurcation stretch has a maximum. The red curves are where mode switches from short modes to long modes.  As $r_2 \to \infty$, the red and black curves intersect at $r_1=0.75$ approximately, after which the red curve coincides with the black curve (so no mode switching takes place for $r_1$ large enough).
The values of $r_2$ at the ends of the dotted vertical lines starting at $a_3=0.004$,  $b_3=0.42$, and $d_3=0.56$ are $0.24$, $1.15$, and $1.21$, respectively.
(b) Dependence of the sign of $c_1$ on $r_1$ and $r_2$ with $G_3$ having coordinates $(0.57, 0.57)$ and the blue line approaching $r_1=0.57$ as $r_2 \to \infty$. The dashed red curves are where mode switching takes place, and the green line coinciding with the red dashed lines is where the sign change of $c_1$ is due to mode jumping, instead of via $c_1=0$ (i.e. $c_1\ne 0$ on the green line except at the two ends). The values of $r_2$ at the ends of the dotted vertical lines starting at $f_3=0.28$ and $g_3=0.53$  are $0.84$ and $1.75$, respectively.}
\label{h10}
\end{center}
\end{figure}

\begin{figure}[h]
\begin{center}
 \includegraphics[width=.99\textwidth]{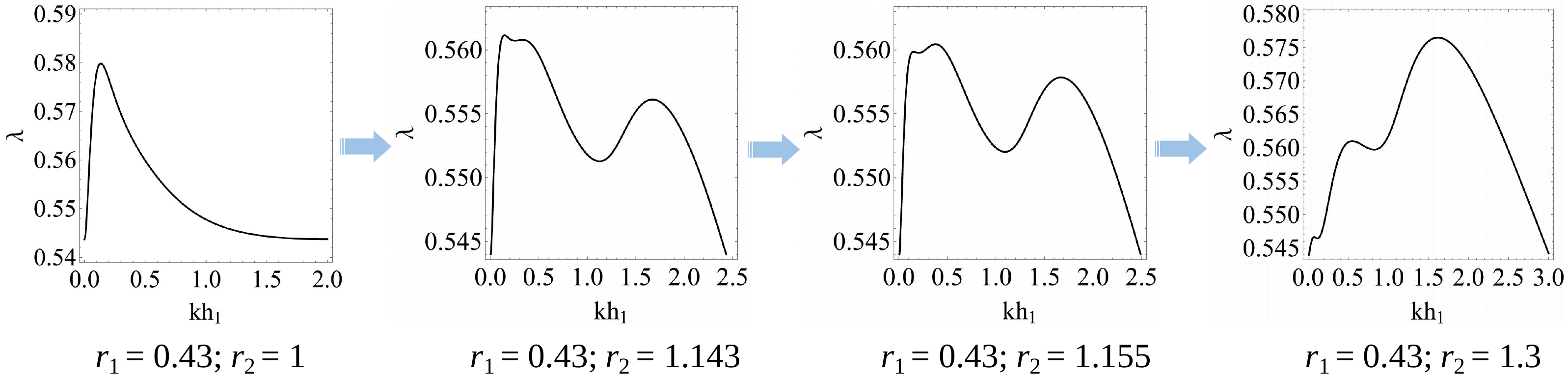}
\caption{Evolution of the bifurcation curve when $h=10,\, r_1=0.43$ showing the fact that as $r_2$ is increased, mode switching may take place among three modes.}
\label{evo3}
\end{center}
\end{figure}

When $h$ is sufficiently larger than $1$ (the lower layer becoming thinner than the top layer), we observe the novel phenomenon that mode switching may take place more than once when $r_2$ is varied while $r_1$ is fixed in a finite interval. This is due to the fact that the bifurcation curve now may have three maxima. Fig.\,\ref{h10}(a) shows the counterpart of Fig.\,\ref{h01}(a) when $h=10$. The red curve is again where mode switching takes place, but now it has two branches. Thus, if $r_1$ is fixed in the interval $(b_3, d_3)$ and $r_2$ is increased gradually from $0$, a long mode is preferred first. This mode gives way to a mode with intermediate wavelength as the short red line is crossed. Finally, this intermediate mode jumps to a short mode as the upper part of the long red line is crossed. An example of this mode switching is displayed in
Fig.\,\ref{evo3} for $r_1=0.43$.

It is clear from the above representative results that the addition of a second layer may provide more flexibility in generating stable periodic patterns. Of course, the existence of a stretch maximum is only one of the necessary conditions for the existence of stable periodic patterns. Another necessary condition is that the associated bifurcations need to be super-critical. This will be discussed in the next section by referring to Figs\,\ref{h1}(b), \ref{h01}(b) and \ref{h10}(b).

\section{Post-buckling analysis}
In this section, we shall first derive the amplitude equation for a single near-critical mode for general prestress and general material models. We then present numerical results for uni-axial compression and for neo-Hookean or Gent materials.

\subsection{General prestress and material models}

For general prestress, we assume that the finitely stressed state $B_{\rm e}$ is determined by a single parameter $\lambda$, which, for instance, in the case of a uni-axial compression is the stretch in the $x_1$-direction. We shall only consider the case when the bifurcation curve has a single maximum for $\lambda$, which we denote by $\lambda_0$. For an analysis of multiple mode interaction, we refer to  \citet{fu1995} and  \citet{Fu-Ogden1999}. We denote by $B_{\rm cr}$ the critical configuration where the stretch in the $x_1$-direction is equal to $\lambda_0$. Guided by the analysis in \citet{cai-fu1999}, we assume that
\begin{equation}
\lambda = \lambda_0 + \epsilon^{2} \lambda_{1}, \phantom{aa}
\bar{p} = \bar{p}_{0} + \epsilon^{2} \lambda_{1} \bar{p}_{1}, \la{ase41}
\end{equation}
where $\lambda_1$ is a constant, and $\bar{p}_{0}$ and $\bar{p}_{1}$ can be expressed in terms of $\lambda_0$ (and also the elastic moduli). We denote the uniformly deformed configuration associated with \rr{ase41} by $B_{\rm e}$. We observe that the expansions \rr{dg22}$-$\rr{dpt21} are valid for any finitely deformed state, e.g. the critical state $(\lambda_0, \bar{p}_0)$ or the perturbed state $(\lambda, \bar{p})$. As an alternative to the approach adopted by \citet{cai-fu1999} where the independent variables $x_1$ and $x_2$ are defined in $B_{\rm e}$, in the current post-buckling analysis these variables are defined in $B_{\rm cr}$. It is then convenient to identify the $\bar{F}$ and  $\bar{p}$ in \rr{dg22}$-$\rr{dpt21} with $ {\rm diag}\{\lambda_0, \lambda_0^{-1}\}$ and $\bar{p}_{0}$, respectively. In other words, we assume that the expansions \rr{dg22}$-$\rr{dpt21} are around the critical configuration $B_{\rm cr}$. We shall show that the current approach and the approach of \citet{cai-fu1999} give the same expression for the nonlinear coefficient.

We look for a solution of the form
\be
u_j =   \ep^2 u^{(0)}_i \lambda_1 + \epsilon u^{(1)}_j (x_1,x_2) + \epsilon^2 u^{(2)}_j (x_1,x_2)+ \epsilon^3 u^{(3)}_j (x_1,x_2) + \cdot \cdot \cdot,  \la{as41} \en
\be
   \tilde{p} =\epsilon^{2} \bar{p}_{1} \lambda_1 + \epsilon p^{(1)} (x_1,x_2) + \epsilon^2 p^{(2)} (x_1,x_2) + \epsilon^3 p^{(3)} (x_1,x_2) + \cdot \cdot \cdot, \la{as42} \en
where the first terms represent the uniform perturbation from $B_{\rm cr}$ to $B_{\rm e}$, and are given by
$ u^{(0)}_1=  x_1/\lambda_0$, $ u^{(0)}_2=-x_2/\lambda_0$,
whereas the leading-order solution takes the form
\be u^{(1)}_1 = \psi_{,2}, \phantom{aaaa}
u^{(1)}_2 = - \psi_{,1},  \;\;\;\; \psi = A  H (x_2) E + {\rm c.c.},  \;\;\;\; E = \textup{e}^{\textup{i} x_1},\la{los41} \en
and
\be p^{(1)}=A  P_1 (x_2) E + {\rm c.c.},\;\;\;\;P_1 (x_2)=-\ii \gamma H'''+\ii ( {\cal A}^1_{1111}-{\cal A}^1_{1122}-{\cal A}^1_{1221}) H'. \la{los42} \en
In the last two expressions, c.c denotes the complex conjugate of the preceding term, $A$ is the unknown amplitude that is to be determined, and $H (x_2)$ is given by \rr{an31}.  The expression for $E$ indicates that we have chosen the mode number of the near-critical mode to be unity. This is without loss of generality since the bifurcation condition takes the form $F(kh_1, h, \lambda, r_1, r_2)=0$ and so we can obtain different values of $k$ by varying $h_1$ alone. Alternatively, for each fixed $h_1$, we may use the inverse of the critical wavenumber to scale $x_1$ and $x_2$, and then the wavenumber becomes unity relative to the scaled coordinates. In the following analysis, the $h_1$ is the thickness of the first layer in the critical configuration $B_{\rm cr}$ and is equal to the critical value of $kh_1$ in the linear analysis.
We have separated the
$O(\ep^2)$ uniform perturbations in \rr{as41} and \rr{as42} from the other $O(\ep^2)$ terms  in order to facilitate comparisons with the expressions in \citet{cai-fu1999}.

Since the above leading-order solution is simply the linearized solution with the undetermined amplitude chosen to be $A$, when we substitute \rr{as41}$-$\rr{as42} into the nonlinear governing equations and auxiliary conditions, we find that the system of equations obtained by equating the coefficients of $\epsilon$ is automatically satisfied. By equating the coefficients of $\epsilon^2$, we obtain the second-order governing equations
\begin{equation}
u^{(2)}_{1,1} + u^{(2)}_{2,2} = \frac{1}{2} u^{(1)}_{i,j} u^{(1)}_{j,i},\la{soge41}
\end{equation}
\begin{equation}
\mathcal{A}^1_{jilk} u^{(2)}_{k,lj} - p^{(2)}_{,i} = - \mathcal{A}^2_{jilknm} u^{(1)}_{m,n} u^{(1)}_{k,lj} - p^{(1)}_{,j} u^{(1)}_{j,i},\la{soge42}
\end{equation}
and the corresponding auxiliary conditions

\begin{equation}
{T}^{(2)}_{i} = 0 \phantom{aa}
\textup{on} \phantom{a} x_2 = h_1+h_2,\la{ac41}
\end{equation}
\begin{equation}
[{u}_i^{(2)}] = 0, \phantom{aaa}
[{T}^{(2)}_{i}] = 0, \phantom{aaa}
\textup{on} \phantom{a} x_2 =0, \; {\rm or}\; h_1,\la{ac42}
\end{equation}
\begin{equation}
{u}_i^{(2)} \to 0 \phantom{aa} \textup{as} \phantom{a} x_2 \to -\infty,\la{ac44}
\end{equation}
where
\begin{equation}
T^{(2)}_i = \mathcal{A}^1_{2ilk} u^{(2)}_{k,l} + \bar{p}_{0} u^{(2)}_{2,i} - p^{(2)} \delta_{2i} + \frac{1}{2} \mathcal{A}^2_{2ilknm} u^{(1)}_{k,l} u^{(1)}_{m,n} - \bar{p}_{0} u^{(1)}_{2,k} u^{(1)}_{k,i} + p^{(1)} u^{(1)}_{2,i}.\la{trn41}
\end{equation}
In writing down the last expression, we have made use of the result $\mathcal{A}^1_{2ilk} u^{(0)}_{k,l} + \bar{p}_{0} u^{(0)}_{2,i} - \bar{p}_1 \delta_{2i} \equiv 0$ which is the traction-free boundary condition associated with the infinitesimal homogeneous perturbation from $B_{\rm e}$ to $B_{\rm cr}$. As a result,
$\lambda_1$ does not appear in \rr{soge41}$-$\rr{trn41} which take the same form as in \citet{cai-fu1999}.

The second-order problem specified by \rr{soge41}$-$\rr{trn41} can be solved once the form of prestress and the elastic moduli are given. In the next section we will explain how this problem can be solved and present explicit results for the special case of uniaxial compression and neo-Hookean materials. However, our following derivation of the amplitude equation does not depend on the explicit solution of this problem.

To derive the amplitude equation which must be satisfied by $A$, we follow \citet{fu1995} and \citet{Fu-Devenish1996} and make use of the virtual work principle
\begin{equation}
\int_{- \infty}^{h_1 + h_2} \textup{d} x_2 \int_{0}^{2 \pi}  \chi_{ij}  u^{*}_{i,j} \textup{d} x_1 = 0,\la{ci44} \en
where  $u^{*}_{i}$ is a linear solution corresponding to $k=-1$. More precisely, we take \be
u^{*}_{1} = {H}^{'} (x_2) \textup{e}^{- \textup{i} x_1}, \phantom{aaa}
u^{*}_{2} = \textup{i} {H} (x_2) \textup{e}^{- \textup{i} x_1}, \;\;\;\; -\infty < x_2 \leq h_1+h_2.
\la{ls43} \en
The identity \rr{ci44} can be proved by integration by part followed by an application of the divergence theorem \citep{cai-fu1999}.
The expansions \rr{as41}$-$\rr{as42} can now be substituted into \rr{tf22} and the resulting expression into \rr{ci44}. On equating the coefficients of $\epsilon$, $\epsilon^2$ and $\epsilon^3$, it is found that the two equations obtained from equating the coefficients of $\epsilon$ and $\epsilon^2$ are automatically satisfied. From equating the coefficient of $\epsilon^3$, we obtain
\be
\int_{- \infty}^{h_1 + h_2} \textup{d} x_2 \int_{0}^{2 \pi} \bigg\{
 ({\cal L}_{ij} [\bm{u}^{(3)},p^{(3)}] + \sigma^{(3)}_{ij}) u^{*}_{i,j} \bigg\} \textup{d} x_1 = 0, \la{ci45}\en
where
\be {\cal L}_{ij} [\bm{u}^{(3)},p^{(3)}] = \mathcal{A}^{1}_{jilk} u^{(3)}_{k,l} + \bar{p}_0 u^{(3)}_{j,i} - p^{(3)} \delta_{ji}, \la{cont41}\en
 $$\sigma^{(3)}_{ij} = \mathcal{A}^{2}_{jilknm} u^{(1)}_{k,l} (u^{(2)}_{m,n}+ \lambda_1 u^{(0)}_{m,n}) + \frac{1}{6} \mathcal{A}^{3}_{jilknmqp} u^{(1)}_{k,l} u^{(1)}_{m,n} u^{(1)}_{p,q} $$ \be + p^{(1)} (u^{(2)}_{j,i}+ \lambda_1 u^{(0)}_{j,i}) + ( \lambda_1 \bar{p}_1+p^{(2)}) u^{(1)}_{j,i}.\la{cont43}\en
In writing down \rr{cont43} we have made of the identities that
$$
(u^{(1)}_{j,k}u^{(2)}_{k,i}+u^{(2)}_{j,k}u^{(1)}_{k,i}-u^{(1)}_{j,k}u^{(1)}_{k,l}u^{(1)}_{l,i}) u^{*}_{i,j}=0,
\;\;\;\; u^{(1)}_{j,k}u^{(1)}_{k,i} u^{*}_{i,j}=0, $$
which can be verified by expanding the summations and then making use of the properties $u^{(1)}_{j,j}=u^{*}_{j,j}=0$ and \rr{soge41}.

Only the first term in the integrand of \rr{ci45} now contains the unknown third-order solution $(\bm{u}^{(3)},p^{(3)})$. By integrating ${\cal L}_{ij} [\bm{u}^{(3)},p^{(3)}] u^{*}_{i,j}$ repeatedly by parts and making use of the fact that $u^{*}_{i}$ given by \rr{ls43} is a linear solution, this term can be expressed in terms of the first- and second-order solutions, and \rr{ci45} then reduces to
\be
\int_{- \infty}^{h_1 + h_2} \textup{d} x_2 \int_{0}^{2 \pi} \bigg\{
  p^{*} u^{(1)}_{j,i} u^{(2)}_{i,j} + \sigma^{(3)}_{ij} u^{*}_{i,j} \bigg\} \textup{d} x_1 = 0,\la{ci46}\en
where $p^{*}$ is the pressure field corresponding to $u^{*}_i$, and is given by
$$p^{*}= \left\{\ii \gamma H'''+\ii ({\cal A}^1_{1122}+{\cal A}^1_{1221}-{\cal A}^1_{1111}) H' \right\} \textup{e}^{- \textup{i} x_1}. $$
In obtaining \rr{ci46}, we have also made use of the result
$
u^{(3)}_{i,i} = u^{(1)}_{i,j}  u^{(2)}_{j,i},
$
which is obtained from equating the coefficients of $\epsilon^{3}$ in \rr{dpt21}.

To facilitate the remaining presentation, we write
\be u^{(1)}_{i,j} = A \varGamma^{(1)}_{ij} E + c.c., \phantom{aa}
u^{*}_{i,j} = \bar{\varGamma}^{(1)}_{ij} \bar{E} , \la{rexp41}\en
\be p^{(1)} = A P_1 E + c.c., \phantom{aa}
p^{(2)} = A \bar{A} P_0 + A^2 P_2 E^2 + c.c., \la{rexp42}\en
\be u^{(2)}_{i,j} = A \bar{A} \varGamma^{(m)}_{ij} +  A^2 \varGamma^{(2)}_{ij} E^2 + c.c., \la{rexp43}\en
where the bars on $A$ and $\varGamma_{ij}$ signify complex conjugation and the expressions for $\varGamma^{(1)}_{ij}$, $\varGamma^{(m)}_{ij}$, $\varGamma^{(2)}_{ij}$, $P_1$, $P_0$, $P_2$ can be obtained from the leading-order and second-order solutions.

On substituting \rr{rexp41}$-$\rr{rexp43} into \rr{ci46} and evaluating the integral with respect to $x_1$, we obtain the amplitude equation
\be
  c_0 \lambda_1 A + c_1 |A|^2 A =0,\la{ee41}
\en
where the linear and nonlinear coefficients $c_0$ and $c_1$ are given, respectively, by
\be
c_0= \int^{h_1 + h_2}_{- \infty} \left\{ \mathcal{A}^{2}_{jilknm} u^{(0)}_{m,n}  \varGamma^{(1)}_{kl} +P_1 u^{(0)}_{j,i} + \bar{p}_1 \varGamma^{(1)}_{ji}   \right\}  \bar{\varGamma}^{(1)}_{ij} \textup{d} x_2, \phantom{aa}
 \la{comt42} \en
\be c_1 = \int^{h_1 + h_2}_{- \infty} \{ \bar{P}_1 (\varGamma^{(1)}_{ij} \varGamma^{(m)}_{ji} + 2 \bar{\varGamma}^{(1)}_{ij} \varGamma^{(2)}_{ji} ) + K_{ij} \bar{\varGamma}^{(1)}_{ij} \} \textup{d} x_2. \la{comt43} \en
In the above expression for $c_1$, the $K_{ij}$ is given by
 $$K_{ij} = \mathcal{A}^{2}_{jilknm} (\varGamma^{(1)}_{kl} \varGamma^{(m)}_{mn}  + \bar{\varGamma}^{(1)}_{kl} \varGamma^{(2)}_{mn}) + \frac{1}{2} \mathcal{A}^{3}_{jilknmqp} \bar{\varGamma}^{(1)}_{kl} \varGamma^{(1)}_{mn} \varGamma^{(1)}_{pq} $$ \be
 + P_1 \varGamma^{(m)}_{ji} + P_0 \varGamma^{(1)}_{ji} + P_2 \bar{\varGamma}^{(1)}_{ji} .\la{kij41}\en
The amplitude equation \rr{ee41} admits the non-trivial post-buckling solution
\begin{equation}
|A|^2 = -\frac{c_0}{c_1}\lambda_1.
\nonumber
\end{equation}
It can be shown \citep{cai-fu1999} that $c_0$ is always positive. Thus the above solution can be obtained only if $\lambda_1 / c_1 < 0$. It then follows that the bifurcation is supercritical if $c_1 > 0$ and subcritical if $c_1 < 0$.

 On comparing the above expressions for $c_0$ and $c_1$ with those in  \citet{cai-fu1999}, we see that the two expressions for $c_1$ are identical, but those for $c_0$ are different. The discrepancy can be explained by noting that if we were to make a variable transformation from the coordinates in $B_{\rm e}$ to $B_{\rm cr}$ in the expression for $\chi_{ij}$ in \citet{cai-fu1999}, the following extra term would be produced to order $\ep^3$:
 \be \ep^3 \frac{\lambda_1}{\lambda_0} \left\{ (-1)^l  \mathcal{A}^{1}_{jilk} u_{k,l}^{(1)} + (-1)^i \bar{p}_0 u_{j,i}^{(1)}\right\}, \;\;\;\;(\hbox{no summation on}\; i \; \hbox{but summation on}\; l ). \la{extra} \en
  We have verified numerically that when this term is added in the evaluation of the virtual work principle in \citet{cai-fu1999}, the approach used by  \citet{cai-fu1999} gives the same result for $c_0$ as the approach adopted in the current paper. We note, however, that this discrepancy is immaterial since it is the sign of $c_1$ that determines whether the bifurcation is super-critical or sub-critical.

\subsection{Uniaxial compression and neo-Hookean materials}
In this subsection we calculate the coefficients in the amplitude equation \rr{ee41} for the special case when the layers and half-space are made of different neo-Hookean materials and the prestress takes the form of a uniaxial compression. We shall present numerical results for the three cases considered in Figs\,\ref{h1}(a),  \ref{h01}(a) and \ref{h10}(a).


We assume that the maximum $\lambda_0$ in \rr{ase41} is attained at $k h_1 = h_{1cr}$. Since we have taken $k=1$, this implies that $h_1 =h_{1cr}$.
We note that both $\lambda_0$ and $h_{1cr}$ depend on $r_1$ and $r_2$, and $\bar{p}_1$ is related to $\lambda_1$ by $\bar{p}_1=-2 \mu \lambda_1/\lambda_0^3$.  Our aim in this subsection is to determine the dependence of $c_1$ on $r_1$ and $r_2$.

With the aid of \rr{los41}$-$\rr{los42} and \rr{rexp41}$_1$, we obtain
\be
\varGamma^{(1)}_{11} = \textup{i} H^{'}, \phantom{aaa}
\varGamma^{(1)}_{12} = H^{''}, \phantom{aaa}
\varGamma^{(1)}_{21} = H, \phantom{aaa}
\varGamma^{(1)}_{22} = - \textup{i} H^{'},\la{f41}
\en
and
\be P_1(x_2) = \textup{i} \mu (\lambda^{2} H^{'} - \lambda^{-2} H^{'''}). \la{p1} \en
The governing equations \rr{soge41} and \rr{soge42} for the second-order solution reduce to
\be
u^{(2)}_{1,1} + u^{(2)}_{2,2} = (u^{(1)}_{1,1})^{2} + u^{(1)}_{1,2} \phantom{.} u^{(1)}_{2,1}, \phantom{aaa}
\mu \bar{B}_{jl} u^{(2)}_{i,jl} - p^{(2)}_{,i} = - p^{(1)}_{,j} u^{(1)}_{j,i}, \la{u41}
\end{equation}
where $\bar{\bm B}={\rm diag} \left\{\lambda_0^2, \lambda_0^{-2}\right\}$.
These equations are to be solved subject to the
auxiliary conditions \rr{ac41}$-$\rr{ac44}, where the expression \rr{trn41} for $T^{(2)}_i$ now reduces to
\begin{equation}
T^{(2)}_i = \mu \bar{B}_{2l} u^{(2)}_{i,l} + \bar{p}_0 u^{(2)}_{2,i} - p^{(2)} \delta_{2i} - \bar{p}_0 u^{(1)}_{2,k} u^{(1)}_{k,i} + p^{(1)} u^{(1)}_{2,i}.
\la{t41}
\end{equation}
Due to quadratic interaction, the right-hand side of \rr{u41}$_{1}$ is a linear combination of $E^{0}$, $E^{2}$ and $E^{-2}$. Thus the solution for $u^{(2)}_i$ takes the form
\be u^{(2)}_1 = A \bar{A} U_0 (x_2) + A^{2} U_2 (x_2) E^2 + c.c., \la{uu41} \en
\be u^{(2)}_2 = A \bar{A} V_0 (x_2) + A^{2} V_2 (x_2) E^2 + c.c., \la{uu42} \en
and $p^{(2)}$ takes the form \rr{rexp42}$_2$.  On substituting \rr{uu41}$-$\rr{uu42} and \rr{rexp42}$_2$ into \rr{u41}, we find, after some manipulation,
\begin{equation}
U_0 = 0, \phantom{aa}
V_0 = 2 H H^{'}, \phantom{aa}
2 \textup{i} U_2 = - V^{'}_2 + H H^{''} - H^{'2},
\la{u0v041}
\end{equation}
\begin{equation}
P_0 = 2 \mu \lambda^{-2} (H H^{'})^{'} + 2 \textup{i} P_1 H^{'}, \phantom{aa}
2 \textup{i} P_2 = \mu \lambda^{-2} U_2^{''} - 4 \mu \lambda^{2} U_2 + P_1^{'} H - P_1 H^{'},
\la{p0p241}
\end{equation}
\begin{equation}
V^{''''}_2 - 4(1 + \lambda^{4}) V^{''}_2 + 16 \lambda^{4} V_2 = 3(\lambda^{4} - 1)(H^{'} H^{''} - H H^{'''}).
\la{vv241}
\end{equation}
It was shown in \citet{cai-fu1999} that \rr{vv241} has a particular integral given by
\be V= \varGamma(x_2)  \equiv \frac{\lambda^{4}-1}{9 \lambda^{8} - 82 \lambda^{4} + 9} (9 H H^{'''} - 21 H^{'} H^{''} ).\la{fx241} \en
Thus, the general solution to \rr{vv241} is given by
\begin{equation}
V_2 = \left\{ \begin{array}{ll} B_1 \phantom{.} \textup{exp}(2 s_1 x_2) + B_2 \phantom{.} \textup{exp}(2 s_2 x_2) + \varGamma(x_2), & x_2 \in (-\infty, 0) \\
\sum_{j=1}^{4} \hat{B}_{j} \phantom{.} \textup{exp} (2 {s}_j x_2) + {\varGamma}(x_2), & x_2 \in (0, h_1) \\
\sum_{j=1}^{4} \tilde{B}_{j} \phantom{.} \textup{exp} (2 {s}_j x_2) + {\varGamma}(x_2), & x_2 \in (h_1, h_1+h_2), \end{array} \right.
\la{v241}
\end{equation}
where $s_1=1, s_2=\lambda_0^2$, $s_3=-1, s_4=-\lambda_0^2$, and $B_1$, $B_2$, $\hat{B}_1$ to $\hat{B}_4$, $\tilde{B}_1$ to $\tilde{B}_4$ are constants that are determined by the auxiliary conditions \rr{ac41}$-$\rr{ac44}.

With the aid of \rr{uu41}$-$\rr{uu42},  we may calculate $u^{(2)}_{i,j}$. Comparing the resulting expressions with \rr{rexp43} then yields
$$
\varGamma^{(m)}_{11} = 0 , \phantom{aaaa}
\varGamma^{(m)}_{12} = U^{'}_0 , \phantom{aaaa}
\varGamma^{(m)}_{21} = 0 , \phantom{aaaa}
\varGamma^{(m)}_{22} = V^{'}_0 ,$$
\be \varGamma^{(2)}_{11} = 2 \textup{i} U_2 , \phantom{aaa}
\varGamma^{(2)}_{12} = U^{'}_2 , \phantom{aaa}
\varGamma^{(2)}_{21} = 2 \textup{i} V^{'}_2  , \phantom{aaa}
\varGamma^{(2)}_{22} = V^{'}_2 .
\la{ff41}
\en
The expressions for $c_0$ and $c_1$ can  be simplified further by noting that the second- and third-order elastic moduli are all zero.

We now investigate the dependence of $c_1$ on $r_1$ and $r_2$ for the three representative cases shown in Figs \ref{h1}(a), \ref{h01}(a) and \ref{h10}(a). As in the linear analysis, some general results may be deduced by referring to the three limits, $r_1 \to 1$, $r_2 \to r_1$, and $r_2 \to \infty$,  under which the structure under consideration reduces to a half-space coated by a single layer. Since for the latter reduced case $c_1$ vanishes when the modulus ratio is equal to $0.57$, we may deduce for the current problem that the curve corresponding to $c_1=0$ must contain the point $(r_1, r_2)=(0.57, 0.57)$ and must approach the asymptote $r_1=0.57$ as $r_2 \to \infty$.
These facts are used to validate the Mathematica code that is used to compute $c_1$ for any choice of $r_1, r_2$ and $h$ for which a stretch maximum exists.

Fig.\,\ref{h1}(b) shows the sign of $c_1$ in the $(r_1, r_2)$-plane when $h=1$. The plane is divided into three regions by two solid curves, and the three regions correspond to $c_1>0$, $c_1<0$, and  non-existence of a stretch maximum, respectively. In addition to the general observations made above, the blue solid curve also tends to an asymptote as $r_1 \to \infty$. Fitting the numerical results for $ 2.1<r_1<10$ to a straight line, we obtain $r_2=0.38 r_1+0.24$ which is displayed in Fig.\,\ref{h1}(b). Our results conform with the expectation that the bifurcation will be supercritical if both layers are much stiffer than the half-space (corresponding to the area near the origin in Fig.\,\ref{h1}(b)), but there are also two novel aspects. Firstly, even if both layers are softer than the half-space (i.e. $r_1>1, r_2>1$), the bifurcation can still be supercritical (and so robust wrinkling patterns can be observed) provided $r_2 <0.38 r_1+0.24$, that is if the top layer is sufficiently harder than the first layer. Secondly, if $r_2$ is fixed to lie in the interval $(0.414, 1)$ and $r_1$ is increased from zero, then $c_1$ changes sign twice: it is positive for sufficiently small or large values of $r_1$, but is negative in between. Similarly, when $r_1$ is fixed to be between $0.57$ and $0.648$ and $r_2$ is increased from zero, the $c_1$ also changes sign twice. Thus, adding an extra layer enables robust wrinkling patterns to be achieved over a larger parameter regime.

\begin{figure}[h]
\begin{center}
\begin{tabular}{ccc}
\includegraphics[scale=0.7]{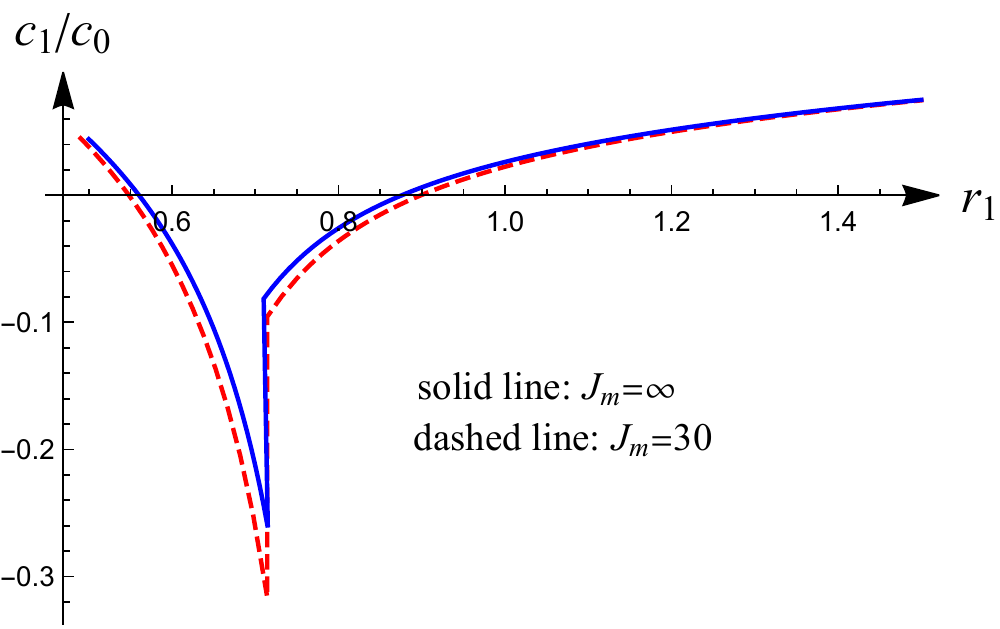}& &\includegraphics[scale=0.7]{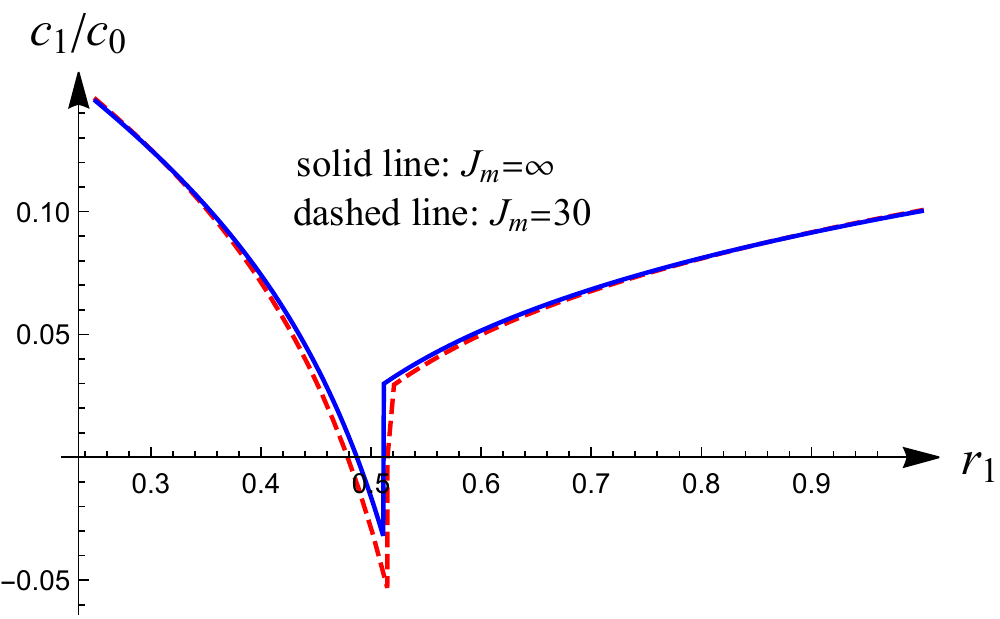}\\
(a) $r_2=0.5, h=0.1$ & & (b) $r_2=0.25, h=0.1$
\end{tabular}
\caption{Comparison of dependence of $c_1/c_0$ on $r_1$ when the Gent material model with $J_m=30$ or the neo-Hookean model is used. Two typical mode jumping behaviors are shown. (a) $c_1$ does not change sign when mode jumping takes place at $r_1=0.72$, $\lambda_0=0.63$, with associated wave numbers given by $(kh_1)_{\rm cr}=1.39,\,14.57$; (b) $c_1$  changes sign when mode jumping takes place at $r_1=0.52$, $\lambda_0=0.70$, with associated wave numbers given by $(kh_1)_{\rm cr}=1.17,\,12.34$.}
\label{fig7}
\end{center}
\end{figure}

Figure \ref{h01}(b) displays the corresponding results when $h=0.1$. The asymptotes associated with the limits $r_1 \to \infty$ and $r_2 \to \infty$ are similar to those in the previous case, but now the curve corresponding to $c_1=0$ splits into two branches due to the presence of mode switching. These two branches are connected by the green line across which the sign of $c_1$ changes abruptly due to mode switching. On the other two segments of the red dashed line across which mode jumping takes place, the sign of $c_1$ remains unchanged when the line is crossed; see later discussion related to Figure \ref{fig7}. Note that the vertical asymptote $r_2 =0.57$ lies between $f_2$ and $g_2$. Thus,
when $r_1$ is fixed to lie in the interval $(f_2, 0.57)$ and $r_2$ is increased from zero gradually, the nature of bifurcation changes according to supercritical $\to$ subcritical $\to$ supercritical, whereas when $r_1$ is fixed to lie in the interval $(0.57, 0.60)$ it evolves like  supercritical $\to$ subcritical $\to$ supercritical $\to$ subcritical.

Finally, in Figure \ref{h10}(b) we display the results for $h=10$. In the limit $r_2 \to \infty$, the black and blue lines asymptote to $r_1=1$ and $r_1=0.57$, respectively, whereas the red dotted line intersect the black line at $r_1= 0.75$ approximately, after which
the red curve stays on the black curve (so no mode switching takes place for $r_1$ large enough).  The curve corresponding to $c_1=0$ again splits into two branches due to the presence of mode switching.  If we now fix $r_1$ and increase $r_2$ from zero gradually, the bifurcation behaviour is again dependent on the fixed value of $r_1$ but even more complicated than in the previous case. In both cases when mode switching is possible, the mode switching lines (the red dashed lines) consist of three distinctive parts: a part that is entirely in the domain of $c_1>0$, a part that is entirely in the domain of $c_1<0$, and a part across which the sign of $c_1$ changes. Since bifurcation with $c_1<0$ is sensitive to imperfections, we expect that mode switching can only take place in a predictable fashion across the first part.

\subsection{Uniaxial compression and general material model}
 When more general strain energy functions are used, solution of the second order problem follows the same procedure as in the previous section. The only major difference is that a particular integral as simple as \rr{fx241} does not seem to be possible. Instead the required particular integral in \rr{v241} can be found using the method of variation of parameters \citep{fu-cai2015}. If the equation for $V_2$ is written in the form
\be
V^{''''}_2 -(1+s^2) a^2 V^{''}_2 + s^2 a^4 V_2 = \omega(x_2), \la{add9}\en
where $a$ and $s$ are known constants, then the particular integral is given by
$$
{\varGamma}(x_2)= \frac{1}{2 a^3 s (1-s^2)} \left\{ s {\rm e}^{  a x_2} \int {\rm e}^{-  a x_2} \omega(x_2) dx_2 -
{\rm e}^{  a s x_2} \int  {\rm e}^{-  a s x_2} \omega(x_2) dx_2 \right. $$ \be \left. + {\rm e}^{-a s x_2} \int {\rm e}^{ a s x_2} \omega(x_2) dx_2 -
s {\rm e}^{- a  x_2} \int {\rm e}^{ a  x_2} \omega(x_2) dx_2  \right\}, \la{add10} \en
with the understanding that the arbitrary constants in the indefinite integrals are all set to zero (this is necessary in order to satisfy the decay condition as $x_2 \to -\infty$).
This expression is valid for the two layers as well as the substrate although  $\omega(x_2)$ takes different expressions in the three different regions.

We have written a separate Mathematica code based on the above formula to compute $c_1$ for any material model.  The programme is used to validate the programme used in the previous subsection which is written specifically and independently for the neo-Hookean material model. As an illustrative example to show the effect of the material extensibility $J_m$, in Fig.\,\ref{fig7}(a, b) we have shown  $c_1/c_0$ against $r_1$ when $h=0.1$ and $r_2$ is fixed to be $0.5$ and $0.25$, respectively. The leading order solution is normalised such that
 $u_2=A \cos x_1$ at $x_2=h_1+h_2$. The results are displayed against their counterparts when the neo-Hookean model is used for which the results are already given in Fig.\,\ref{h01}(b). It is seen that decreasing $J_m$ has little effect on the value of $r_2$ at which mode jumping takes place and only slightly widens the interval of $r_1$ where $c_1$ is negative.

\section{Conclusion}
In this paper we have investigated the linear and nonlinear buckling properties of a hyperelastic half-space coated with two layers. At macro-scales, buckling usually undermines a structure's integrity and should be avoided. When a structure is sensitive to imperfections, any imperfection, material or geometrical, will significantly reduce the critical load at which bifurcation takes place. Thus from a practical point of view, it is important to find the parameter regime in which the structure is imperfection sensitive. At micrometer and sub-micrometer scales, robust wrinkling patterns can be harnessed to serve useful purposes. Since only supercritical bifurcations may be observable/realizable in practice, results from our weakly nonlinear analysis provide a road map on how to choose a variety of combinations of material parameters to achieve robust wrinkling patterns.

 Our analysis is conducted with the aid of the exact theory of nonlinear elasticity and for general strain energy functions. A Mathematica code is written for computing the coefficient $c_1$ the sign of which determines whether the bifurcation is supercritical or not.  For the current two-layers/substrate structure, $c_1$ depends on the modulus ratios $r_1$ and $r_2$ as well as the thickness ratio $h$. For each fixed $h$, we may display the sign of $c_1$ in the
$(r_1, r_2)$-plane, covering all the possibilities. Illustrative results are presented for the case when the material is modelled by the neo-Hookean model or the Gent model, and the prestress takes the form of a uniaxial compression.

When the neo-Hookean model is used, three sets of representative results are presented corresponding to $h=0.1, 1$ and $10$, respectively. They illustrate the three possibilities of no mode switching (when $h=1$), mode switching occurring once (when $h=0.1$), and mode switching occurring twice (when $h=10$), respectively.
For each case, we display in the $(r_1, r_2)$-plane domains where the stretch has a maximum and where $c_1$ is positive. One important finding is that when mode switching is theoretically possible based on the linear analysis, it may not be observable/realisable/controllable if it occurs on a part of the mode switching line where $c_1$ is negative or changes sign. When the Gent model is used, we determine the effects of varying the extensibility parameter $J_m$ and it is found that changing $J_m$ does not seem to change our results in any qualitative way, and the quantitative differences it makes are still insignificant when $J_m$ has become as small as $30$.

We remark that although we have presented some representative behaviours, our numerical calculations are by no means intended to be exhaustive. For instance, we cannot conclude whether three or more stretch maxima can occur or not for other parameter combinations. Neither have we considered the effects of allowing a pre-stretch in the substrate \citep{hutchinson2013}. The main aim of this paper has been to demonstrate that the sign of $c_1$ can be computed semi-analytically, with the aid of Mathematica, without making any approximations even for the most general material model.  Our Mathematica code is freely available to any interested reader upon request.
\subsection*{Acknowledgements}
This work was supported by the National Natural Science Foundation of China (Grant Nos 11672202).

\begin{thebibliography}{61}
\expandafter\ifx\csname natexlab\endcsname\relax\def\natexlab#1{#1}\fi
\providecommand{\url}[1]{\texttt{#1}}
\providecommand{\href}[2]{#2}
\providecommand{\path}[1]{#1}
\providecommand{\DOIprefix}{doi:}
\providecommand{\ArXivprefix}{arXiv:}
\providecommand{\URLprefix}{URL: }
\providecommand{\Pubmedprefix}{pmid:}
\providecommand{\doi}[1]{\href{http://dx.doi.org/#1}{\path{#1}}}
\providecommand{\Pubmed}[1]{\href{pmid:#1}{\path{#1}}}
\providecommand{\bibinfo}[2]{#2}
\ifx\xfnm\relax \def\xfnm[#1]{\unskip,\space#1}\fi
\bibitem[{Alawiye et~al.(2020)Alawiye, Farrell \& Goriely}]{Alawiye2020}
\bibinfo{author}{Alawiye, H.}, \bibinfo{author}{Farrell, E.}, \&
  \bibinfo{author}{Goriely, A.} (\bibinfo{year}{2020}).
\newblock \bibinfo{title}{Revisiting the wrinkling of elastic bilayers ii:
  post-bifurcation analysis}.
\newblock {\it \bibinfo{journal}{J. Mech. Phys. Solids}\/},  {\it
  \bibinfo{volume}{143}\/}, \bibinfo{pages}{104053}.
\bibitem[{Alawiye et~al.(2019)Alawiye, Kuhl \& Goriely}]{Alawiye2019}
\bibinfo{author}{Alawiye, H.}, \bibinfo{author}{Kuhl, E.}, \&
  \bibinfo{author}{Goriely, A.} (\bibinfo{year}{2019}).
\newblock \bibinfo{title}{Revisiting the wrinkling of elastic bilayers i:
  linear analysis}.
\newblock {\it \bibinfo{journal}{Phil. Tran. R. Soc. A}\/},  {\it
  \bibinfo{volume}{377}\/}, \bibinfo{pages}{20180076}.
\bibitem[{Audoly \& Boudaoud(2008)}]{AB2008}
\bibinfo{author}{Audoly, B.}, \& \bibinfo{author}{Boudaoud, A.}
  (\bibinfo{year}{2008}).
\newblock \bibinfo{title}{Buckling of a stiff film bound to a compliant
  substrate -- part i: formulation, linear stability of cylindrical patterns,
  secondary bifurcations}.
\newblock {\it \bibinfo{journal}{J. Mech. Phys. Solids}\/},  {\it
  \bibinfo{volume}{56}\/}, \bibinfo{pages}{2401--2421}.
\bibitem[{Bigoni et~al.(1997)Bigoni, Ortiz \& Needleman}]{bon1997}
\bibinfo{author}{Bigoni, D.}, \bibinfo{author}{Ortiz, M.}, \&
  \bibinfo{author}{Needleman, A.} (\bibinfo{year}{1997}).
\newblock \bibinfo{title}{Effect of interfacial compliance on bifurcation of a
  layer bonded to a substrate}.
\newblock {\it \bibinfo{journal}{Int. J. Solids Struct.}\/},  {\it
  \bibinfo{volume}{34}\/}, \bibinfo{pages}{4305--4326}.
\bibitem[{Biot(1963)}]{bi1963}
\bibinfo{author}{Biot, M.~A.} (\bibinfo{year}{1963}).
\newblock \bibinfo{title}{Surface instability of rubber in compression}.
\newblock {\it \bibinfo{journal}{Appl. Sci. Res. Sect. A}\/},  {\it
  \bibinfo{volume}{12}\/}, \bibinfo{pages}{168--182}.
\bibitem[{Bowden et~al.(1998)Bowden, Brittain, Evans, Hutchinson \&
  Whitesides}]{boden1998}
\bibinfo{author}{Bowden, N.}, \bibinfo{author}{Brittain, S.},
  \bibinfo{author}{Evans, A.~G.}, \bibinfo{author}{Hutchinson, J.~W.}, \&
  \bibinfo{author}{Whitesides, G.~M.} (\bibinfo{year}{1998}).
\newblock \bibinfo{title}{Spontaneous formation of ordered structures in thin
  films of metals supported on an elastomeric polymer}.
\newblock {\it \bibinfo{journal}{Nature}\/},  {\it \bibinfo{volume}{393}\/},
  \bibinfo{pages}{146--149}.
\bibitem[{Bowden et~al.(1999)Bowden, Huck, E \& Whitesides}]{boden1999}
\bibinfo{author}{Bowden, N.}, \bibinfo{author}{Huck, W. T.~S.},
  \bibinfo{author}{E, P.~K.}, \& \bibinfo{author}{Whitesides, G.~M.}
  (\bibinfo{year}{1999}).
\newblock \bibinfo{title}{The controlled formation of ordered, sinusoidal
  structures by plasma oxidation of an elastomeric polymer}.
\newblock {\it \bibinfo{journal}{Appl. Phys. Lett.}\/},  {\it
  \bibinfo{volume}{75}\/}, \bibinfo{pages}{2557--2559}.
\bibitem[{Brau et~al.(2011)Brau, Vandeparre, Sabbah, Poulard, Boudaoud \&
  Damman}]{brau2011}
\bibinfo{author}{Brau, F.}, \bibinfo{author}{Vandeparre, H.},
  \bibinfo{author}{Sabbah, A.}, \bibinfo{author}{Poulard, C.},
  \bibinfo{author}{Boudaoud, A.}, \& \bibinfo{author}{Damman, P.}
  (\bibinfo{year}{2011}).
\newblock \bibinfo{title}{Multiple-length-scale elastic instability mimics
  parametric resonance of nonlinear oscillators}.
\newblock {\it \bibinfo{journal}{Nat. Phys}\/},  {\it \bibinfo{volume}{7}\/},
  \bibinfo{pages}{56--60}.
\bibitem[{Budday et~al.(2015)Budday, Kuhl \& Hutchinson}]{budday2015}
\bibinfo{author}{Budday, S.}, \bibinfo{author}{Kuhl, E.}, \&
  \bibinfo{author}{Hutchinson, J.~W.} (\bibinfo{year}{2015}).
\newblock \bibinfo{title}{Period-doubling and period-tripling in growing
  bilayered systems}.
\newblock {\it \bibinfo{journal}{Phil. Mag.}\/},  {\it \bibinfo{volume}{95}\/},
  \bibinfo{pages}{3208--3309}.
\bibitem[{Cai et~al.(2012)Cai, Chen, Suo \& Hayward}]{cai2012}
\bibinfo{author}{Cai, S.~Q.}, \bibinfo{author}{Chen, D.~Y.},
  \bibinfo{author}{Suo, Z.~G.}, \& \bibinfo{author}{Hayward, R.~C.}
  (\bibinfo{year}{2012}).
\newblock \bibinfo{title}{Creasing instability of elastomer films}.
\newblock {\it \bibinfo{journal}{Soft Matter}\/},  {\it \bibinfo{volume}{8}\/},
  \bibinfo{pages}{1301--1304}.
\bibitem[{Cai \& Fu(1999)}]{cai-fu1999}
\bibinfo{author}{Cai, Z.~X.}, \& \bibinfo{author}{Fu, Y.~B.}
  (\bibinfo{year}{1999}).
\newblock \bibinfo{title}{On the imperfection sensitivity of a coated elastic
  half-space}.
\newblock {\it \bibinfo{journal}{Proc. R. Soc. Lond. A}\/},  {\it
  \bibinfo{volume}{455}\/}, \bibinfo{pages}{3285--3309}.
\bibitem[{Cai \& Fu(2000)}]{cai-fu2000}
\bibinfo{author}{Cai, Z.~X.}, \& \bibinfo{author}{Fu, Y.~B.}
  (\bibinfo{year}{2000}).
\newblock \bibinfo{title}{Exact and asymptotic stability analyses of a coated
  elastic half-space}.
\newblock {\it \bibinfo{journal}{Int. J. Solids Struct.}\/},  {\it
  \bibinfo{volume}{37}\/}, \bibinfo{pages}{3101--3119}.
\bibitem[{Cai \& Fu(2019)}]{cai-fu2019}
\bibinfo{author}{Cai, Z.~X.}, \& \bibinfo{author}{Fu, Y.~B.}
  (\bibinfo{year}{2019}).
\newblock \bibinfo{title}{Effects of pre-stretch, compressibility and material
  constitution on the period-doubling secondary bifurcation of a film/substrate
  bilayer}.
\newblock {\it \bibinfo{journal}{Int. J. Non-linear Mech.}\/},  {\it
  \bibinfo{volume}{115}\/}, \bibinfo{pages}{11--19}.
\bibitem[{Cao \& Hutchinson(2012)}]{ch2012a}
\bibinfo{author}{Cao, Y.~P.}, \& \bibinfo{author}{Hutchinson, J.~W.}
  (\bibinfo{year}{2012}).
\newblock \bibinfo{title}{Wrinkling phenomena in neo-hookean film/substrate
  bilayers}.
\newblock {\it \bibinfo{journal}{ASME J. Appl. Mech.}\/},  {\it
  \bibinfo{volume}{79}\/}, \bibinfo{pages}{031019}.
\bibitem[{Chadwick \& Ogden(1971)}]{co1971}
\bibinfo{author}{Chadwick, P.}, \& \bibinfo{author}{Ogden, R.~W.}
  (\bibinfo{year}{1971}).
\newblock \bibinfo{title}{On the definition of elastic moduli}.
\newblock {\it \bibinfo{journal}{Arch. Ration. Mech. Anal.}\/},  {\it
  \bibinfo{volume}{44}\/}, \bibinfo{pages}{41--53}.
\bibitem[{Chan et~al.(2008)Chan, Smith, Hayward \& Crosby}]{chan2008}
\bibinfo{author}{Chan, E.~P.}, \bibinfo{author}{Smith, E.~J.},
  \bibinfo{author}{Hayward, R.~C.}, \& \bibinfo{author}{Crosby, A.~J.}
  (\bibinfo{year}{2008}).
\newblock \bibinfo{title}{Surface wrinkles for smart adhesion}.
\newblock {\it \bibinfo{journal}{Adv. Mater.}\/},  {\it
  \bibinfo{volume}{20}\/}, \bibinfo{pages}{711--716}.
\bibitem[{Chen \& Hutchinson(2004)}]{CH2004}
\bibinfo{author}{Chen, X.}, \& \bibinfo{author}{Hutchinson, J.~W.}
  (\bibinfo{year}{2004}).
\newblock \bibinfo{title}{Herring bone buckling patterns of compressed thin
  films on compliant substrates}.
\newblock {\it \bibinfo{journal}{J. Appl. Mech.}\/},  {\it
  \bibinfo{volume}{71}\/}, \bibinfo{pages}{597--603}.
\bibitem[{Cheng et~al.(2014)Cheng, Zhang, Hwang, Rogers \& Huang}]{czh2014}
\bibinfo{author}{Cheng, H.~Y.}, \bibinfo{author}{Zhang, Y.~H.},
  \bibinfo{author}{Hwang, K.~C.}, \bibinfo{author}{Rogers, J.~A.}, \&
  \bibinfo{author}{Huang, Y.~G.} (\bibinfo{year}{2014}).
\newblock \bibinfo{title}{Buckling of a stiff thin film on a pre-strained
  bi-layer substrate}.
\newblock {\it \bibinfo{journal}{Int. J. Solids Struct.}\/},  {\it
  \bibinfo{volume}{51}\/}, \bibinfo{pages}{3113--3118}.
\bibitem[{Cheng \& Xu(2020)}]{cx2020}
\bibinfo{author}{Cheng, Z.}, \& \bibinfo{author}{Xu, F.}
  (\bibinfo{year}{2020}).
\newblock \bibinfo{title}{Intricate evolutions of multiple-period post-buckling
  patterns in bilayers}.
\newblock {\it \bibinfo{journal}{Science China Physics, Mechanics \&
  Astronomy}\/},  (pp. \bibinfo{pages}{doi.org/10.1007/s11433--020--1620--0}).
\bibitem[{Chien et~al.(2012)Chien, Kuo, Wang, Tsai \& Tsai}]{chien2012}
\bibinfo{author}{Chien, H.-W.}, \bibinfo{author}{Kuo, W.-H.},
  \bibinfo{author}{Wang, M.-J.}, \bibinfo{author}{Tsai, S.-W.}, \&
  \bibinfo{author}{Tsai, W.-B.} (\bibinfo{year}{2012}).
\newblock \bibinfo{title}{Tunable micropatterned substrates based on
  poly(dopamine) deposition via microcontact printing}.
\newblock {\it \bibinfo{journal}{Langmuir}\/},  {\it \bibinfo{volume}{28}\/},
  \bibinfo{pages}{5775--5782}.
\bibitem[{Ciarletta(2014)}]{Ci2014}
\bibinfo{author}{Ciarletta, P.} (\bibinfo{year}{2014}).
\newblock \bibinfo{title}{Wrinkle-to-fold transition in soft layers under
  equi-biaxial strain: a weakly nonlinear analysis}.
\newblock {\it \bibinfo{journal}{J. Mech. Phys. Solids}\/},  {\it
  \bibinfo{volume}{73}\/}, \bibinfo{pages}{118--133}.
\bibitem[{Ciarletta \& Fu(2015)}]{CF2015}
\bibinfo{author}{Ciarletta, P.}, \& \bibinfo{author}{Fu, Y.~B.}
  (\bibinfo{year}{2015}).
\newblock \bibinfo{title}{A semi-analytical approach to biot instability in a
  growing layer: Strain gradient correction, weakly non-linear analysis and
  imperfection sensitivity}.
\newblock {\it \bibinfo{journal}{Int. J. Non-linear Mech.}\/},  {\it
  \bibinfo{volume}{75}\/}, \bibinfo{pages}{38--45}.
\bibitem[{Dimmock et~al.(2020)Dimmock, Wang, Fu, El~Haj \& Yang}]{dw2020}
\bibinfo{author}{Dimmock, R.~L.}, \bibinfo{author}{Wang, X.~L.},
  \bibinfo{author}{Fu, Y.}, \bibinfo{author}{El~Haj, A.~J.}, \&
  \bibinfo{author}{Yang, Y.} (\bibinfo{year}{2020}).
\newblock \bibinfo{title}{Biomedical applications of wrinkling polymers}.
\newblock {\it \bibinfo{journal}{Recent progress in materials}\/},  {\it
  \bibinfo{volume}{2}\/}, \bibinfo{pages}{2001005}.
\bibitem[{Dorris \& Nemat-Nasser(1980)}]{dn1980}
\bibinfo{author}{Dorris, J.~F.}, \& \bibinfo{author}{Nemat-Nasser, S.}
  (\bibinfo{year}{1980}).
\newblock \bibinfo{title}{Instability of a layer on a half-space}.
\newblock {\it \bibinfo{journal}{J. Appl. Mech.}\/},  {\it
  \bibinfo{volume}{47}\/}, \bibinfo{pages}{304--312}.
\bibitem[{Dowaikh \& Ogden(1991)}]{do1991}
\bibinfo{author}{Dowaikh, M.~A.}, \& \bibinfo{author}{Ogden, R.~W.}
  (\bibinfo{year}{1991}).
\newblock \bibinfo{title}{Interfacial waves and deformations in pre-stressed
  elastic media}.
\newblock {\it \bibinfo{journal}{Proc. Roy. Soc.}\/},  {\it
  \bibinfo{volume}{433}\/}, \bibinfo{pages}{313--328}.
\bibitem[{Fu(1995)}]{fu1995}
\bibinfo{author}{Fu, Y.~B.} (\bibinfo{year}{1995}).
\newblock \bibinfo{title}{Resonant-triad instability of a pre-stressed
  incompressible elastic plate}.
\newblock {\it \bibinfo{journal}{J. Elast.}\/},  {\it \bibinfo{volume}{41}\/},
  \bibinfo{pages}{13--37}.
\bibitem[{Fu(2005)}]{fu2005}
\bibinfo{author}{Fu, Y.~B.} (\bibinfo{year}{2005}).
\newblock \bibinfo{title}{An explicit expression for the surface-impedance
  matrix of a generally anisotropic incompressible elastic material in a state
  of plane strain}.
\newblock {\it \bibinfo{journal}{Int. J. Non-Linear Mech.}\/},  {\it
  \bibinfo{volume}{40}\/}, \bibinfo{pages}{229--239}.
\bibitem[{Fu \& Cai(2015)}]{fu-cai2015}
\bibinfo{author}{Fu, Y.~B.}, \& \bibinfo{author}{Cai, Z.~X.}
  (\bibinfo{year}{2015}).
\newblock \bibinfo{title}{An asymptotic analysis of the period-doubling
  secondary bifurcation in a film/substrate bilayer}.
\newblock {\it \bibinfo{journal}{SIAM J. Appl. Math.}\/},  {\it
  \bibinfo{volume}{75}\/}, \bibinfo{pages}{2381--2395}.
\bibitem[{Fu \& Ciarletta(2014)}]{FC2014}
\bibinfo{author}{Fu, Y.~B.}, \& \bibinfo{author}{Ciarletta, P.}
  (\bibinfo{year}{2014}).
\newblock \bibinfo{title}{Buckling of a coated elastic half-space when the
  coating and substrate have similar material properties}.
\newblock {\it \bibinfo{journal}{Proc. R. Soc. Lond. A}\/},  {\it
  \bibinfo{volume}{471}\/}, \bibinfo{pages}{20140979}.
\bibitem[{Fu \& Devenish(1996)}]{Fu-Devenish1996}
\bibinfo{author}{Fu, Y.~B.}, \& \bibinfo{author}{Devenish, B.}
  (\bibinfo{year}{1996}).
\newblock \bibinfo{title}{Effects of pre-stresses on the propagation of
  nonlinear surface waves in an elastic half-space}.
\newblock {\it \bibinfo{journal}{Q. Jl Mech. Appl. Math.}\/},  {\it
  \bibinfo{volume}{49}\/}, \bibinfo{pages}{65--80}.
\bibitem[{Fu \& Ogden(1999)}]{Fu-Ogden1999}
\bibinfo{author}{Fu, Y.~B.}, \& \bibinfo{author}{Ogden, R.~W.}
  (\bibinfo{year}{1999}).
\newblock \bibinfo{title}{Nonlinear stability analysis of pre-stressed elastic
  bodies}.
\newblock {\it \bibinfo{journal}{Continuum Mech. Thermodynam.}\/},  {\it
  \bibinfo{volume}{11}\/}, \bibinfo{pages}{141--172}.
\bibitem[{Fu \& Rogerson(1994)}]{fr1994}
\bibinfo{author}{Fu, Y.~B.}, \& \bibinfo{author}{Rogerson, G.~A.}
  (\bibinfo{year}{1994}).
\newblock \bibinfo{title}{A nonlinear analysis of instability of a pre-stressed
  incompressible elastic platet}.
\newblock {\it \bibinfo{journal}{Proc. R. Soc. Lond.}\/},  {\it
  \bibinfo{volume}{446}\/}, \bibinfo{pages}{233--254}.
\bibitem[{Genzer \& Groenewold(2006)}]{gg2006}
\bibinfo{author}{Genzer, J.}, \& \bibinfo{author}{Groenewold, J.}
  (\bibinfo{year}{2006}).
\newblock \bibinfo{title}{Soft matter with hard skin: from skin wrinkles to
  templating and material characterization}.
\newblock {\it \bibinfo{journal}{Soft Matter}\/},  {\it \bibinfo{volume}{2}\/},
  \bibinfo{pages}{310--323}.
\bibitem[{Goriely(2017)}]{goriely2017}
\bibinfo{author}{Goriely, A.} (\bibinfo{year}{2017}).
\newblock {\it \bibinfo{title}{The mathematics and mechanics of biological
  growth}\/}.
\newblock \bibinfo{publisher}{Springer}.
\bibitem[{Holland et~al.(2017)Holland, Li, Feng \& Kuhl}]{holland2017}
\bibinfo{author}{Holland, M.~A.}, \bibinfo{author}{Li, B.},
  \bibinfo{author}{Feng, X.~Q.}, \& \bibinfo{author}{Kuhl, E.}
  (\bibinfo{year}{2017}).
\newblock \bibinfo{title}{Instabilities of soft films on compliant substrates}.
\newblock {\it \bibinfo{journal}{J. Mech. Phy. Solids}\/},  {\it
  \bibinfo{volume}{98}\/}, \bibinfo{pages}{350--365}.
\bibitem[{Huang et~al.(2005)Huang, Hong \& Suo}]{HH2005}
\bibinfo{author}{Huang, Z.~Y.}, \bibinfo{author}{Hong, W.}, \&
  \bibinfo{author}{Suo, Z.} (\bibinfo{year}{2005}).
\newblock \bibinfo{title}{Nonlinear analyses of wrinkles in a film bonded to a
  compliant substrate}.
\newblock {\it \bibinfo{journal}{J. Mech. Phys. Solids}\/},  {\it
  \bibinfo{volume}{53}\/}, \bibinfo{pages}{2101--2118}.
\bibitem[{Hutchinson(2013)}]{hutchinson2013}
\bibinfo{author}{Hutchinson, J.~W.} (\bibinfo{year}{2013}).
\newblock \bibinfo{title}{The role of nonlinar substrate elasticity in the
  wrinkling of thin films}.
\newblock {\it \bibinfo{journal}{Phil. Trans. R. Soc., A}\/},  {\it
  \bibinfo{volume}{371}\/}, \bibinfo{pages}{20120422}.
\bibitem[{Jia et~al.(2012)Jia, Cao, Liu, Jiang, Feng \& Yu}]{jcl2012}
\bibinfo{author}{Jia, F.}, \bibinfo{author}{Cao, Y.~P.}, \bibinfo{author}{Liu,
  T.~S.}, \bibinfo{author}{Jiang, Y.}, \bibinfo{author}{Feng, X.~Q.}, \&
  \bibinfo{author}{Yu, S.~W.} (\bibinfo{year}{2012}).
\newblock \bibinfo{title}{Wrinkling of a bilayer resting on a soft substrate
  under in-plane compression}.
\newblock {\it \bibinfo{journal}{Phil. Mag.}\/},  {\it \bibinfo{volume}{92}\/},
  \bibinfo{pages}{1554--1568}.
\bibitem[{Kim et~al.(2013)Kim, Hu, Alvarenga, Kolle, Suo \&
  Aizenberg}]{kim2013}
\bibinfo{author}{Kim, P.}, \bibinfo{author}{Hu, Y.},
  \bibinfo{author}{Alvarenga, J.}, \bibinfo{author}{Kolle, M.},
  \bibinfo{author}{Suo, Z.}, \& \bibinfo{author}{Aizenberg, J.}
  (\bibinfo{year}{2013}).
\newblock \bibinfo{title}{Rational design of mechano-responsive optical
  materials by fine tuning the evolution of strain-dependent wrinkling
  patterns}.
\newblock {\it \bibinfo{journal}{Adv. Opt. Mater.}\/},  {\it
  \bibinfo{volume}{1}\/}, \bibinfo{pages}{381--388}.
\bibitem[{Lee et~al.(2010)Lee, Lee, Lim, Lee, Lee \& Cho}]{lee2010}
\bibinfo{author}{Lee, S.~G.}, \bibinfo{author}{Lee, D.~Y.},
  \bibinfo{author}{Lim, H.~S.}, \bibinfo{author}{Lee, D.~H.},
  \bibinfo{author}{Lee, S.}, \& \bibinfo{author}{Cho, K.}
  (\bibinfo{year}{2010}).
\newblock \bibinfo{title}{Switchable transparency and wetting of elastomeric
  smart windows}.
\newblock {\it \bibinfo{journal}{Adv. Mater.}\/},  {\it
  \bibinfo{volume}{22}\/}, \bibinfo{pages}{5013--5017}.
\bibitem[{Lejeune et~al.(2016)Lejeune, Javili \& Christian}]{ljc2016}
\bibinfo{author}{Lejeune, E.}, \bibinfo{author}{Javili, A.}, \&
  \bibinfo{author}{Christian, L.} (\bibinfo{year}{2016}).
\newblock \bibinfo{title}{Understanding geometric instabilities in thin films
  via a multi-layer model}.
\newblock {\it \bibinfo{journal}{Soft Matter}\/},  {\it
  \bibinfo{volume}{12}\/}, \bibinfo{pages}{806--816}.
\bibitem[{Li et~al.(2012)Li, Cao, Feng \& Gao}]{li-cao2012}
\bibinfo{author}{Li, B.}, \bibinfo{author}{Cao, Y.~P.}, \bibinfo{author}{Feng,
  X.~Q.}, \& \bibinfo{author}{Gao, H.~J.} (\bibinfo{year}{2012}).
\newblock \bibinfo{title}{Mechanics of morphological instabilities and surface
  wrinkling in soft materials: a review}.
\newblock {\it \bibinfo{journal}{Soft Matter}\/},  {\it \bibinfo{volume}{8}\/},
  \bibinfo{pages}{5728--5745}.
\bibitem[{Liu \& Bertoldi(2015)}]{liu2015}
\bibinfo{author}{Liu, J.}, \& \bibinfo{author}{Bertoldi, K.}
  (\bibinfo{year}{2015}).
\newblock \bibinfo{title}{Bloch wave approach for the analysis of sequential
  bifurcations in bilayer structures}.
\newblock {\it \bibinfo{journal}{Proc. R. Soc. Lond. A}\/},  {\it
  \bibinfo{volume}{471}\/}, \bibinfo{pages}{20150493}.
\bibitem[{Ma et~al.(2013)Ma, Liang, Chen, Poon, Jiang \& H}]{ma2013}
\bibinfo{author}{Ma, T.}, \bibinfo{author}{Liang, H.}, \bibinfo{author}{Chen,
  G.}, \bibinfo{author}{Poon, B.}, \bibinfo{author}{Jiang, H.}, \&
  \bibinfo{author}{H, Y.} (\bibinfo{year}{2013}).
\newblock \bibinfo{title}{Micro-strain sensing using wrinkled stiff thin films
  on soft substrates as tunable optical grating}.
\newblock {\it \bibinfo{journal}{Opt. Express}\/},  {\it
  \bibinfo{volume}{21}\/}, \bibinfo{pages}{11994--12001}.
\bibitem[{Nolte et~al.(2006)Nolte, Cohen \& Rubner}]{ncr2006}
\bibinfo{author}{Nolte, A.~J.}, \bibinfo{author}{Cohen, R.~E.}, \&
  \bibinfo{author}{Rubner, M.~F.} (\bibinfo{year}{2006}).
\newblock \bibinfo{title}{A two-plate buckling technique for thin film modulus
  measurements:? applications to polyelectrolyte multilayers}.
\newblock {\it \bibinfo{journal}{Macromolecules}\/},  {\it
  \bibinfo{volume}{39}\/}, \bibinfo{pages}{4841--4847}.
\bibitem[{Ogden \& Sotiropoulos(1996)}]{os1996}
\bibinfo{author}{Ogden, R.~W.}, \& \bibinfo{author}{Sotiropoulos, D.~A.}
  (\bibinfo{year}{1996}).
\newblock \bibinfo{title}{The effect of pres-stress on guided ultrasonic waves
  between a surface layer and a half-space}.
\newblock {\it \bibinfo{journal}{Ultrasonics}\/},  {\it
  \bibinfo{volume}{34}\/}, \bibinfo{pages}{491--494}.
\bibitem[{Rambausek \& Danas(2021)}]{rd2021}
\bibinfo{author}{Rambausek, M.}, \& \bibinfo{author}{Danas, K.}
  (\bibinfo{year}{2021}).
\newblock \bibinfo{title}{Bifurcation of magnetorheological film-substrate
  elastomers subjected to biaxial pre-compression and transverse magnetic
  fields}.
\newblock {\it \bibinfo{journal}{Int. J. Non-linear Mech.}\/},  {\it
  \bibinfo{volume}{128}\/}, \bibinfo{pages}{103608}.
\bibitem[{Shield et~al.(1994)Shield, Kim \& Shield}]{sks1994}
\bibinfo{author}{Shield, T.~W.}, \bibinfo{author}{Kim, K.~S.}, \&
  \bibinfo{author}{Shield, R.~T.} (\bibinfo{year}{1994}).
\newblock \bibinfo{title}{The buckling of an elastic layer bonded to an elastic
  substrate in plane strain}.
\newblock {\it \bibinfo{journal}{J. Appl. Mech.}\/},  {\it
  \bibinfo{volume}{61}\/}, \bibinfo{pages}{231--235}.
\bibitem[{Song et~al.(2008)Song, Jiang, Liu, Khang, Huang, Rogers, Lu \&
  Koh}]{SJ2008}
\bibinfo{author}{Song, J.}, \bibinfo{author}{Jiang, H.}, \bibinfo{author}{Liu,
  Z.~J.}, \bibinfo{author}{Khang, D.~Y.}, \bibinfo{author}{Huang, Y.},
  \bibinfo{author}{Rogers, J.~A.}, \bibinfo{author}{Lu, C.}, \&
  \bibinfo{author}{Koh, C.~G.} (\bibinfo{year}{2008}).
\newblock \bibinfo{title}{Buckling of a stiff thin film on a compliant
  substrate in large deformation}.
\newblock {\it \bibinfo{journal}{Int. J. Solids Struct.}\/},  {\it
  \bibinfo{volume}{45}\/}, \bibinfo{pages}{3107--3121}.
\bibitem[{Stafford et~al.(2004)Stafford, Harrison, Beers, Karim, J,
  VanLandingham, Kim, Volksen, Miller \& Simonyi}]{stafford2004}
\bibinfo{author}{Stafford, C.~M.}, \bibinfo{author}{Harrison, C.},
  \bibinfo{author}{Beers, K.~L.}, \bibinfo{author}{Karim, A.},
  \bibinfo{author}{J, A.~E.}, \bibinfo{author}{VanLandingham, M.~R.},
  \bibinfo{author}{Kim, H.~C.}, \bibinfo{author}{Volksen, W.},
  \bibinfo{author}{Miller, R.~D.}, \& \bibinfo{author}{Simonyi, E.~E.}
  (\bibinfo{year}{2004}).
\newblock \bibinfo{title}{A buckling-based metrology for measuring the elastic
  moduli of polymeric thin films}.
\newblock {\it \bibinfo{journal}{Nat. Mater.}\/},  {\it \bibinfo{volume}{3}\/},
  \bibinfo{pages}{545--550}.
\bibitem[{Steigmann \& Ogden(1997)}]{so1997}
\bibinfo{author}{Steigmann, D.~J.}, \& \bibinfo{author}{Ogden, R.~W.}
  (\bibinfo{year}{1997}).
\newblock \bibinfo{title}{Plane deformations of elastic solids with intrinsic
  boundary elasticity}.
\newblock {\it \bibinfo{journal}{Proc. R. Soc. Lond. A}\/},  {\it
  \bibinfo{volume}{453}\/}, \bibinfo{pages}{853--877}.
\bibitem[{Steigmann \& Ogden(2002)}]{so2002}
\bibinfo{author}{Steigmann, D.~J.}, \& \bibinfo{author}{Ogden, R.~W.}
  (\bibinfo{year}{2002}).
\newblock \bibinfo{title}{Plane strain dynamics of elastic solids with
  intrinsic boundary elasticity, with application to surface wave propagation}.
\newblock {\it \bibinfo{journal}{J. Mech. Phys. Solids}\/},  {\it
  \bibinfo{volume}{50}\/}, \bibinfo{pages}{1869--1896}.
\bibitem[{Sun et~al.(2012)Sun, Xia, M-Y, H \& K-S}]{sx2012}
\bibinfo{author}{Sun, J.-Y.}, \bibinfo{author}{Xia, S.}, \bibinfo{author}{M-Y,
  M.}, \bibinfo{author}{H, O.~K.}, \& \bibinfo{author}{K-S, K.}
  (\bibinfo{year}{2012}).
\newblock \bibinfo{title}{Folding wrinkles of a thin stiff layer on a soft
  substrate}.
\newblock {\it \bibinfo{journal}{Proc. R. Soc. A}\/},  {\it
  \bibinfo{volume}{468}\/}, \bibinfo{pages}{932--953}.
\bibitem[{Wang et~al.(2020)Wang, Zhang, Nie, Su, Chen \& Song}]{wzn2020}
\bibinfo{author}{Wang, C.~J.}, \bibinfo{author}{Zhang, S.},
  \bibinfo{author}{Nie, S.}, \bibinfo{author}{Su, Y.~P.},
  \bibinfo{author}{Chen, W.~Q.}, \& \bibinfo{author}{Song, J.~Z.}
  (\bibinfo{year}{2020}).
\newblock \bibinfo{title}{Buckling of a stiff thin film on a bi-layer compliant
  substrate of finite thickness}.
\newblock {\it \bibinfo{journal}{Int. J. Solids Struct.}\/},  {\it
  \bibinfo{volume}{188-189}\/}, \bibinfo{pages}{133--140}.
\bibitem[{Wolfram-Research \& Inc.(2019)}]{wolf2019}
\bibinfo{author}{Wolfram-Research}, \& \bibinfo{author}{Inc.}
  (\bibinfo{year}{2019}).
\newblock {\it \bibinfo{title}{Mathematica: version 12.}\/}.
\newblock \bibinfo{publisher}{Wolfram Research Inc, Champaign, IL}.
\bibitem[{Yang et~al.(2010)Yang, Khare \& Lin}]{yang2010}
\bibinfo{author}{Yang, S.}, \bibinfo{author}{Khare, K.}, \&
  \bibinfo{author}{Lin, P.~C.} (\bibinfo{year}{2010}).
\newblock \bibinfo{title}{Harnessing surface wrinkle patterns in soft matter}.
\newblock {\it \bibinfo{journal}{Adv. Funct. Mater}\/},  {\it
  \bibinfo{volume}{20}\/}, \bibinfo{pages}{2550--2564}.
\bibitem[{Zhang(2017)}]{zhang2017}
\bibinfo{author}{Zhang, T.} (\bibinfo{year}{2017}).
\newblock \bibinfo{title}{Symplectic analysis for wrinkles: A case study of
  layered neo-hookean structures}.
\newblock {\it \bibinfo{journal}{J. Appl. Mech.}\/},  {\it
  \bibinfo{volume}{84}\/}, \bibinfo{pages}{071002}.
\bibitem[{Zhang et~al.(2012)Zhang, Zhang, Zhang, Kim,  \& Gao}]{zhang-gao2012}
\bibinfo{author}{Zhang, Z.~Q.}, \bibinfo{author}{Zhang, T.},
  \bibinfo{author}{Zhang, Y.~W.}, \bibinfo{author}{Kim, K.-S.}, , \&
  \bibinfo{author}{Gao, H.~J.} (\bibinfo{year}{2012}).
\newblock \bibinfo{title}{Strain-controlled switching of hierarchically
  wrinkled surfaces between superhydrophobicity and superhydrophilicity}.
\newblock {\it \bibinfo{journal}{Langmuir}\/},  {\it \bibinfo{volume}{28}\/},
  \bibinfo{pages}{2753--2760}.
\bibitem[{Zhao et~al.(2015)Zhao, Cao, Hong, Wadee \& Feng}]{zhao-cao2015}
\bibinfo{author}{Zhao, Y.}, \bibinfo{author}{Cao, Y.~P.},
  \bibinfo{author}{Hong, W.}, \bibinfo{author}{Wadee, M.~K.}, \&
  \bibinfo{author}{Feng, X.~Q.} (\bibinfo{year}{2015}).
\newblock \bibinfo{title}{Towards a quantitative understanding of
  period-doubling wrinkling patterns occurring in film/substrate bilayer
  systems}.
\newblock {\it \bibinfo{journal}{Proc. R. Soc. Lond. A}\/},  {\it
  \bibinfo{volume}{471}\/}, \bibinfo{pages}{20140695}.
\bibitem[{Zhuo \& Zhang(2015{\natexlab{a}})}]{zhuo-zhang2015b}
\bibinfo{author}{Zhuo, L.~J.}, \& \bibinfo{author}{Zhang, Y.}
  (\bibinfo{year}{2015}{\natexlab{a}}).
\newblock \bibinfo{title}{From period-doubling to folding in stiff film/soft
  substrate system: The role of substrate nonlinearity}.
\newblock {\it \bibinfo{journal}{Int. J. Non-linear Mech.}\/},  {\it
  \bibinfo{volume}{76}\/}, \bibinfo{pages}{1--7}.
\bibitem[{Zhuo \& Zhang(2015{\natexlab{b}})}]{zhuo-zhang2015a}
\bibinfo{author}{Zhuo, L.~J.}, \& \bibinfo{author}{Zhang, Y.}
  (\bibinfo{year}{2015}{\natexlab{b}}).
\newblock \bibinfo{title}{The mode-coupling of a stiff film/compliant substrate
  system in the post-buckling range}.
\newblock {\it \bibinfo{journal}{Int. J. Solids Struct.}\/},  {\it
  \bibinfo{volume}{53}\/}, \bibinfo{pages}{28--37}.

\end{thebibliography}


\end{document}